\documentclass[pre,aps,preprint,amsmath,amssymb]{revtex4}
\usepackage{graphicx}
\usepackage{epsfig}
\usepackage{epstopdf}
\include{graphics}

\DeclareMathOperator{\sech}{sech}

\begin{document}

\title{Universality of the amplitude shift in fast two-pulse collisions in weakly perturbed linear physical systems}

\author{Quan M. Nguyen$^{1}$, Toan T. Huynh$^{2,3}$, and Avner Peleg$^{4}$}

\affiliation{$^{1}$Department of Mathematics, International University, 
Vietnam National University-HCMC, Ho Chi Minh City, Vietnam}

\affiliation{$^{2}$Department of Mathematics, University of Science, 
Vietnam National University-HCMC, Ho Chi Minh City, Vietnam}

\affiliation{$^{3}$
Department of Mathematics, University of Medicine and Pharmacy at Ho Chi Minh City, 
Ho Chi Minh City, Vietnam}

\affiliation{$^{4}$Department of Exact Sciences, Afeka College of Engineering, 
Tel Aviv 69988, Israel}

\date{\today}

\begin{abstract}
We demonstrate that the amplitude shifts in fast two-pulse collisions 
in perturbed linear physical systems with weak nonlinear dissipation 
exhibit universal soliton-like behavior. 
The behavior is demonstrated for linear optical waveguides 
with weak cubic loss and for systems described by 
linear diffusion-advection models with weak quadratic loss. 
We show that in both systems, the expressions for the collision-induced 
amplitude shifts due to the nonlinear loss have the same form as 
the expression obtained for a fast collision between two 
solitons of the nonlinear Schr\"odinger equation in the presence of weak cubic loss. 
Furthermore, we show that the expressions for the amplitude shifts are 
universal in the sense that they are independent  
of the exact details of the initial pulse shapes. 
We demonstrate the universal soliton-like behavior of the collision-induced amplitude 
shifts by carrying out numerical simulations with the two perturbed 
coupled linear evolution models with three different initial conditions corresponding to 
pulses with exponentially decreasing tails, pulses with power-law decreasing tails,  
and pulses that are initially nonsmooth and that develop significant tails during the collision.   
In all six cases we observe very good agreement between 
the analytic predictions for the amplitude shifts and the results of 
the numerical simulations.  
\end{abstract}
% \pacs{42.65.Tg, 05.45.Yv, 47.35.Fg}
\maketitle

% The text of the paper starts here. 

\section{Introduction}
\label{Introduction}
The stable shape preserving pulse solutions of nonlinear wave models,  
which are known as solitons, appear in a variety of fields, including 
optics \cite{Agrawal2001,Mollenauer2006}, condensed matter physics \cite{Malomed89}, 
hydrodynamics \cite{Zakharov84,Newell85}, and plasma physics \cite{Horton96}. 
One of the main properties characterizing solitons is their robustness in soliton collisions, 
that is, the fact that the solitons do not change their shape in the collisions.   
Another important property of solitons, which is manifested in fast inter-soliton collisions, 
is the simple form of the scaling relations for collision-induced changes of soliton parameters, 
such as position, phase, amplitude, and frequency \cite{fast_collisions}.  
This property holds both in the absence of perturbations and in the presence 
of weak perturbations to the integrable nonlinear wave model. 
Consider for example fast collisions between two solitons of the cubic  
nonlinear Schr\"odinger (NLS) equation, which is one of the most widely used 
nonlinear wave models in physics \cite{Malomed89,Zakharov84,Newell85}. 
In the absence of perturbations, the collision-induced changes 
in the phase and position of soliton 1, for example, 
scale as $\eta_{2}/|\Delta\beta|$ and $-\eta_{2}/\Delta\beta^{2}$, 
where $\eta_{j}$ with $j=1,2$ are the soliton amplitudes, $\Delta\beta=\beta_{2}-\beta_{1}$, 
and $\beta_{j}$ with $j=1,2$ are the soliton frequencies \cite{Zakharov84,MM98,CP2005}. 
Furthermore, during fast collisions between NLS solitons in the presence of 
a weak perturbation due to cubic loss, the solitons experience 
amplitude and frequency shifts, which scale as 
$-\epsilon_{3}\eta_{1}\eta_{2}/|\Delta\beta|$ and 
$-\epsilon_{3}\eta_{1}^{2}\eta_{2}/\Delta\beta^{2}$ for soliton 1, 
where $\epsilon_{3}$ is the cubic loss coefficient \cite{PNC2010}. 
Similar simple scaling relations hold for fast two-pulse collisions of NLS solitons in 
the presence of other weak perturbations, such as delayed Raman response 
\cite{Chi89,Malomed91,Kumar98,CP2005,P2004,NP2010}, 
and higher-order nonlinear loss \cite{PC2012}.

The simple form of the scaling relations for collision-induced changes of soliton parameters 
can be attributed to the shape preserving and stability properties of the 
solitons \cite{PNC2010,CP2005,PC2012}. The latter two properties 
are related with the integrability of the nonlinear wave model.
Therefore, one might also relate the simple form of the scaling relations 
for changes in soliton parameters in fast two-soliton collisions to the integrability of the model.
One might expect a very different behavior for collisions between pulses 
that are not shape preserving, since in this case, it is expected that changes 
in pulse shape or instability would lead to the breakdown 
of the simple dynamics observed in fast two-soliton collisions. 
This expectation is especially typical for linear physical systems 
that are weakly perturbed by nonlinear dissipation, since the pulses of the 
linear systems are in general not shape preserving \cite{Tkach97,Agrawal2001,Agrawal89a}. 
However, in Ref. \cite{PNH2017B}, we showed that the opposite might in fact be true. 
That is, we demonstrated that the amplitude shifts in fast two-pulse collisions 
in linear physical systems, weakly perturbed by nonlinear dissipation, exhibit soliton-like scaling behavior. 
The behavior was demonstrated for collisions between Gaussian pulses 
of the following central physical systems: (1) linear optical waveguides with weak cubic loss; 
(2) systems described by linear diffusion-advection models with weak quadratic loss. 
We showed that in both cases, the expressions for the amplitude shifts in fast collisions 
between two Gaussian pulses have the same form as the expression 
for the amplitude shift in a fast collision between two solitons 
of the cubic NLS equation in the presence of weak cubic loss. 
The analytic predictions were confirmed by numerical simulations 
with the corresponding perturbed coupled linear evolution models.

The study in Ref. \cite{PNH2017B} was limited to fast collisions 
between Gaussian pulses. Therefore, based on the results of Ref. \cite{PNH2017B},   
it is unclear if the soliton-like behavior of the collision-induced amplitude shift 
is universal in the sense that it does not depend on the details of the 
initial pulse shapes. In the current paper we address this important question. 
More specifically, we show that the simple soliton-like form of the expressions 
for the collision-induced amplitude shifts is universal in the sense that it 
is independent of the exact details of the initial pulse shapes. 
This is done for both linear optical waveguides with weak cubic loss and for 
systems described by linear diffusion-advection models with weak quadratic loss.       
We explain the universal soliton-like form of the expressions for the 
collision-induced amplitude shifts by noting that changes 
in pulse shapes occurring during a collision due to the effects of dispersion or diffusion 
can be neglected for fast collisions, 
and by noting the conservation of the total energies (or total masses) of the pulses  
by the unperturbed linear evolution models.   
Furthermore, we demonstrate the universal behavior of the amplitude 
shifts by carrying out numerical simulations with the two perturbed 
coupled linear evolution models with three different initial conditions corresponding to 
pulses with exponentially decreasing tails, pulses with power-law decreasing tails, 
and pulses that are initially nonsmooth. We find very good agreement between 
the analytic predictions for the amplitude shifts and the results of 
the numerical simulations in all six cases.  
Surprisingly, the analytic predictions hold even for collisions between  
pulses with initially nonsmooth shapes in linear optical waveguides despite 
of the fast generation of significant pulse tails in this case. 
We explain the good agreement between the analytic and numerical results in the latter 
case by noting that during fast collisions most of the pulse energies 
are still contained in the main bodies of the pulses, and by noting the conservation 
of the total energies of the two pulses by the unperturbed linear propagation model.

The rest of the paper is organized as follows. 
In Sec. \ref{waveguides}, we obtain the expression for the collision-induced amplitude 
shift in a fast two-pulse collision in a linear optical waveguide with weak linear and cubic loss. 
We show that this expression is universal in the sense that it is independent of the details of the initial pulse shapes. 
We then present a comparison of the analytic expression with results 
of numerical simulations with the perturbed coupled linear propagation model 
for three major types of pulses. In Sec. \ref{diffusion}, we obtain the expression for the amplitude 
shift in a fast collision between two concentration pulses in systems described by 
perturbed coupled linear diffusion-advection models with weak linear and quadratic loss. 
We then show that this expression is universal.  
Furthermore, we compare the analytic expression for the amplitude shift 
with results of numerical simulations with the perturbed coupled linear diffusion-advection model 
for three main types of pulses.  
Section \ref{conclusions} is devoted to our conclusions. 
In Appendix \ref{appendA}, we derive the relations between the collision-induced 
amplitude shifts and the collision-induced changes in pulse shapes. 
A description of the procedures used for calculating the values of the collision-induced amplitude shift from 
the analytic expressions and from results of numerical simulations is provided in Appendix \ref{appendB}.

\section{Fast collisions in linear optical waveguides}
\label{waveguides} 
\subsection{Propagation model and initial pulse shapes}
\label{waveguides_model} 
We consider the dynamics of fast collisions between two pulses 
of light in linear optical waveguides with weak linear and cubic loss. 
The dynamics of the collision can be described by the following 
system of perturbed coupled linear propagation equations \cite{PNH2017B,Agrawal2007a,PNC2010}: 
\begin{eqnarray}&&
\!\!\!\!\!\!\!
i\partial_z\psi_{1}\!-\!\mbox{sgn}(\tilde\beta_{2})\partial_{t}^{2}\psi_{1}\!=\!
-i\epsilon_{1}\psi_{1}\!
-\!i\epsilon_{3}|\psi_{1}|^2\psi_{1}\!
-\!2i\epsilon_{3}|\psi_{2}|^2\psi_{1},
\nonumber \\&&
\!\!\!\!\!\!\!
i\partial_z\psi_{2}+id_{1}\partial_{t}\psi _{2}
-\mbox{sgn}(\tilde\beta_{2})\partial_{t}^2\psi_{2}=
-i\epsilon_{1}\psi_{2}-i\epsilon_{3}|\psi_{2}|^2\psi_{2}
\nonumber \\&&
-2i\epsilon_{3}|\psi_{1}|^2\psi_{2},  
\!\!\!\!\!\!\!\!
\label{coll1}
\end{eqnarray}           
where $\psi_{1}$ and $\psi_{2}$ are the envelopes of the electric fields of the pulses, 
$z$ is propagation distance, and $t$ is time \cite{Dimensions1}. 
In Eq. (\ref{coll1}), $d_{1}$ is the group velocity coefficient, 
$\tilde\beta_{2}$ is the second-order dispersion coefficient, and 
$\epsilon_{1}$ and $\epsilon_{3}$ are the linear and cubic loss coefficients, 
which satisfy $0<\epsilon_{1} \ll 1$ and $0<\epsilon_{3} \ll 1$.   
The terms $-\mbox{sgn}(\tilde\beta_{2})\partial_{t}^{2}\psi_{j}$ 
on the left hand side of Eq. (\ref{coll1}) are 
due to the effects of second-order dispersion, 
while $id_{1}\partial_{t}\psi _{2}$ is associated with the group velocity difference. 
The first terms on the right hand side of Eq. (\ref{coll1}) describe linear loss effects, 
while the second and third terms describe intra-pulse and inter-pulse effects due to cubic loss.  
Note that the perturbed coupled propagation model (\ref{coll1}) is based on the assumption 
that the effects of cubic (Kerr) nonlinearity can be neglected. 
This assumption was successfully used in previous experimental and 
theoretical works, see e.g., Refs. \cite{Perry97,Liang2005,Cohen2005b,Cohen2004}. 
In addition, it is assumed that the nonlinear loss is weak, 
and as a result, the effects of higher-order loss 
are also neglected. We emphasize, however, that the effects of higher-order 
loss on the collision-induced amplitude shift can be calculated in a manner 
similar to the one described in Sec. \ref{waveguides_delta_A} 
(see also, Ref. \cite{QMN2018}, where the calculation was carried out for collisions 
between Gaussian pulses).

We are interested in demonstrating universal behavior of the amplitude 
shift in fast two-pulse collisions in the sense that the amplitude shift 
is not very sensitive to the exact details of the pulse shape. 
We therefore consider fast collisions between pulses with generic 
initial pulse shapes and with tails that decay sufficiently fast, 
such that the values of the integrals $\int_{-\infty}^{\infty} dt |\psi_{j}(t,0)|^{2}$ 
are finite. We assume that the pulses can be characterized by 
initial amplitudes $A_{j}(0)$, initial widths $W_{j0}$, initial positions $y_{j0}$, 
and initial phases $\alpha_{j0}$.  
To illustrate the universal behavior of the collision-induced amplitude 
shift we consider three prototypical initial pulse shapes, which represent 
three major types of behavior of the pulse tails before and during the collisions. 
More specifically, we consider the following three types of pulses: 
(1) pulses with exponentially decreasing tails, (2) pulses with power-law decreasing 
tails, (3) pulses that are initially nonsmooth and that develop significant tails 
during the propagation. For concreteness, we demonstrate the universal behavior 
of the amplitude shift using the following representative initial pulses: 
hyperbolic secant pulses in (1), generalized Cauchy-Lorentz pulses in (2), 
and square pulses in (3). The initial envelopes of the electric fields for these 
pulses are given by
\begin{eqnarray} &&
\psi_{j}(t,0)=A_{j}(0)\sech\left[(t-y_{j0})/W_{j0}\right]\exp(i\alpha_{j0}),  
\label{IC1}
\end{eqnarray}
with $j=1,2$ for hyperbolic secant pulses, by 
\begin{eqnarray} &&
\psi_{j}(t,0)=\frac{A_{j}(0)\exp(i\alpha_{j0})}
{1+ 2 \left[(t-y_{j0})/W_{j0} \right]^{4}},
\label{IC2}
\end{eqnarray}
with $j=1,2$ for generalized Cauchy-Lorentz pulses, and by 
\begin{eqnarray} &&
\!\!\!\!\!\!\!\!\!
\psi_{j}(t,0)=
\left\{\begin{array}{l l}
A_{j}(0)\exp(i\alpha_{j0}) & \;\; \mbox{for} \;\;  |t-y_{j0}| \le W_{j0}/2,\\
0 & \;\; \mbox{for} \;\; |t-y_{j0}| > W_{j0}/2,\\
\end{array} \right. 
\label{IC3}
\end{eqnarray}     
with $j=1,2$ for square pulses. We emphasize, however, that similar behavior of the 
collision-induced amplitude shift is observed for other choices of the initial 
pulse shapes.

\subsection{Calculation of the collision-induced amplitude shift}        
\label{waveguides_delta_A} 
The expressions for the collision-induced amplitude shifts are obtained 
under the assumption of a complete fast two-pulse collision. 
The complete collision assumption means that the two pulses 
are well separated at $z=0$ and at the final propagation distance $z=z_{f}$.  
To explain the implications of the fast collision assumption, 
we define the collision length $\Delta z_{c}$, 
which is the distance along which the envelopes of the colliding pulses overlap, 
by $\Delta z_{c}=W_{0}/|d_{1}|$, where for simplicity 
we assume $W_{10}=W_{20}=W_{0}$. The fast collision assumption then means that 
$\Delta z_{c}$ is the shortest length scale in the problem. 
In particular, $\Delta z_{c}\ll z_{D}$, where $z_{D}=W_{0}^{2}/2$ 
is the length scale characterizing the effects of second-order dispersion 
on single-pulse propagation (the dispersion length). 
Using the definitions of $\Delta z_{c}$ and $z_{D}$, 
we obtain $W_{0}|d_{1}|/2 \gg 1$, as the condition for a fast two-pulse collision.

We now show that the condition $W_{0}|d_{1}|/2 \gg 1$ 
is clearly satisfied for collisions in massive multichannel weakly perturbed 
linear optical waveguide systems with tens or hundreds of frequency channels. 
Consider as an example a multichannel linear optical fiber system 
with a total wavelength difference 
$\Delta\lambda=\lambda_{2} - \lambda_{1}=0.045$ $\mu\mbox{m}$, 
where $\lambda_{1}=1.549$ $\mu\mbox{m}$ is the wavelength for the 
highest frequency channel and $\lambda_{2}=1.594$ $\mu\mbox{m}$  
is the wavelength for the lowest frequency channel. 
These values are identical to the ones used in the multichannel optical fiber transmission experiment 
with 109 channels reported in Ref. \cite{Mollenauer2003}.  
The group refractive index values for these wavelengths are $n_{g1}=1.4626$ 
and $n_{g2}=1.4629$, respectively \cite{Agrawal2001,Malitson65,Tan98}.  
Assuming in addition that the pulse width is 20 ps and that $|\tilde\beta_{2}|=1$ 
$\mbox{ps}^2/\mbox{km}$, 
which are typical values for mulitchannel transmission at 10 Gb/s per channel, 
we obtain $W_{0}|d_{1}|/2 = 2 \times 10^{4} \gg 1$ for collisions   
between pulses from the two outermost frequency channels. 
Moreover, since the frequency difference between adjacent channels 
in multichannel transmission is constant, we obtain that in a system with 
109 channels $W_{0}|d_{1}|/2 \simeq 185.2 \gg 1$ for 
collisions between pulses from adjacent channels. 
Therefore, the condition for a fast collision is satisfied for all 
collisions in this linear optical fiber transmission system.     
As a second example, consider linear multichannel transmission in a silicon waveguide 
with $\Delta\lambda=\lambda_{2} - \lambda_{1}=0.05$ $\mu\mbox{m}$, 
$\lambda_{1}=0.75$ $\mu\mbox{m}$, and $\lambda_{2}=0.8$ $\mu\mbox{m}$. 
The group refractive index values for these wavelengths are 
$n_{g1}=4.3597$ and $n_{g2}=4.3285$, respectively \cite{Aspnes83}.  
Assuming that the pulse width is 20 ps and that $|\tilde\beta_{2}|=1$ $\mbox{ps}^2/\mbox{km}$, 
we find $W_{0}|d_{1}|/2 = 2.08 \times 10^{6} \gg 1$ for collisions   
between pulses from the two outermost frequency channels. 
Furthermore, we obtain that in a multichannel system with 109 channels 
$W_{0}|d_{1}|/2 \simeq 1.926 \times 10^{4} \gg 1$ for 
collisions between pulses from adjacent frequency channels.  
Thus, the condition for a fast collision is satisfied for all 
collisions in this linear silicon waveguide transmission system.

The perturbation technique for calculating the collision-induced amplitude shift was 
first presented in our work in Ref. \cite{PNH2017B}. We present here a review 
of this perturbation technique along with some important features that were not discussed 
in Ref. \cite{PNH2017B}. Our perturbation procedure is   
a generalization of the perturbative technique, developed in Refs. 
\cite{PCG2003,PCG2004} for calculating the effects of weak perturbations 
on fast two-soliton collisions. Following the perturbative calculation for 
the two-soliton collision, we look for a solution of Eq. (\ref{coll1}) 
in the form 
\begin{eqnarray}&&
\!\!\!\!\!\!\!
\psi_{j}(t,z)=\psi_{j0}(t,z)+\phi_{j}(t,z), 
\label{coll2}
\end{eqnarray}       
where $j=1,2$, $\psi_{j0}$ are the solutions of Eq. (\ref{coll1}) without 
the inter-pulse interaction terms, and $\phi_{j}$ describe corrections to 
$\psi_{j0}$ due to inter-pulse interaction. By definition, 
$\psi_{10}$ and $\psi_{20}$ satisfy the following two weakly perturbed 
linear propagation equations  
\begin{eqnarray}&&
\!\!\!\!\!\!\!
i\partial_z\psi_{10}\!-\!\mbox{sgn}(\tilde\beta_{2})\partial_{t}^{2}\psi_{10}\!=\!
-i\epsilon_{1}\psi_{10}\!
-\!i\epsilon_{3}|\psi_{10}|^2\psi_{10},
\!\!\!\!\!\!\!\!
\label{coll2_add1}
\end{eqnarray}         
and
\begin{eqnarray}&&
\!\!\!\!\!\!\!
i\partial_z\psi_{20}+id_{1}\partial_{t}\psi _{20}
-\mbox{sgn}(\tilde\beta_{2})\partial_{t}^2\psi_{20}=
\nonumber \\&&
-i\epsilon_{1}\psi_{20}
-i\epsilon_{3}|\psi_{20}|^2\psi_{20}.
%\!\!\!\!\!\!\!\!
\label{coll2_add2}
\end{eqnarray}         
We now substitute the ansatz (\ref{coll2}) into Eq. (\ref{coll1}) 
and use Eqs. (\ref{coll2_add1}) and  (\ref{coll2_add2}) 
to obtain equations for the $\phi_{j}$. We focus attention  
on the calculation of $\phi_{1}$, as the calculation of $\phi_{2}$ is similar. 
Taking into account only leading-order effects of the collision, 
i.e., effects of order $\epsilon_{3}$, we can neglect terms 
containing $\phi_{j}$ on the right hand side of the resulting equation. 
We therefore obtain: 
\begin{equation}
i\partial_z\phi_{1}-\mbox{sgn}(\tilde\beta_{2})\partial_{t}^{2}\phi_{1}=
-2i\epsilon_{3}|\psi_{20}|^2\psi_{10}.
\label{coll3} 
\end{equation}      
We continue to follow the perturbation procedure for fast two-soliton collisions and  
substitute $\psi_{j0}(t,z)=\Psi_{j0}(t,z)\exp[i\chi_{j0}(t,z)]$ and 
$\phi_{1}(t,z)=\Phi_{1}(t,z)\exp[i\chi_{10}(t,z)]$  into Eq. (\ref{coll3}), 
where $\Psi_{j0}$ and $\chi_{j0}$ are real-valued.  
We arrive at the following equation for $\Phi_{1}$: 
\begin{eqnarray} &&
i\partial_{z}\Phi_{1} - \left(\partial_{z}\chi_{10}\right)\Phi_{1}
-\mbox{sgn}(\tilde\beta_{2})\left[ 
\partial_{t}^{2}\Phi_{1}
+2i\left(\partial_{t}\chi_{10}\right)\partial_{t}\Phi_{1}
\right. 
\nonumber \\&&
\left. 
+i\left(\partial_{t}^{2}\chi_{10}\right)\Phi_{1}
-\left(\partial_{t}\chi_{10}\right)^2\Phi_{1}
\right] =
-2i\epsilon_{3}\Psi_{20}^{2}\Psi_{10}.
\label{coll3_add1}
\end{eqnarray} 
The term on the right hand side  of Eq. (\ref{coll3_add1}) is of order $\epsilon_{3}$. 
In addition, since the collision length $\Delta z_{c}$ is of order 
$1/|d_{1}|$, the term $i\partial_{z}\Phi_{1}$ is of order 
$|d_{1}| \times O(\Phi_{1})$. Equating the orders of $i\partial_{z}\Phi_{1}$ 
and $-2\epsilon_{3}\Psi_{20}^{2}\Psi_{10}$, we find that $\Phi_{1}$ 
is of order $\epsilon_{3}/|d_{1}|$. In addition, we observe that 
all other terms on the left hand side of Eq. (\ref{coll3_add1}) are 
of order $\epsilon_{3}/|d_{1}|$ or higher, and can therefore be neglected. 
As a result, the equation for $\Phi_{1}$ in the leading order 
of the perturbative calculation is:  
\begin{eqnarray} &&
\partial_{z}\Phi_{1}=
-2\epsilon_{3}\Psi_{20}^{2}\Psi_{10}.
\label{coll4}
\end{eqnarray} 
Equation (\ref{coll4}) has the same form as the equation obtained 
for a fast collision between two solitons of the NLS equation  
in the presence of weak cubic loss (see Eq. (9) in Ref. \cite{PNC2010}).     
We also note that since the $\Psi_{j0}$ are real-valued, 
$\Phi_{1}$ is real-valued as well.

We calculate the net collision-induced amplitude shift of pulse 1 from  
the net collision-induced change in $\Phi_{1}$.  
For this purpose, we denote by $z_{c}$ the collision distance, 
which is the distance at which the maxima of $|\psi_{j}(t,z)|$ coincide.  
In a fast collision, the collision takes place in a small 
interval $[z_{c}-\Delta z_{c},z_{c}+\Delta z_{c}]$ around $z_{c}$.  
Therefore, the net collision-induced change in the envelope of pulse 1 
$\Delta\Phi_{1}(t,z_{c})$ can be evaluated by: 
$\Delta\Phi_{1}(t,z_{c})=\Phi_{1}(t,z_{c}+
\Delta z_{c})-\Phi_{1}(t,z_{c}-\Delta z_{c})$. 
To calculate $\Delta\Phi_{1}(t,z_{c})$,    
we introduce the approximation $\Psi_{j0}(t,z)=A_{j}(z)\tilde\Psi_{j0}(t,z)$, 
where $\tilde\psi_{j0}(t,z)=\tilde\Psi_{j0}(t,z)\exp[i\chi_{j0}(t,z)]$ is the solution 
of the unperturbed linear propagation equation with unit amplitude.  
We then substitute the approximate expressions for $\Psi_{j0}$ into Eq. (\ref{coll4}) 
and integrate with respect to $z$ over the interval $[z_{c}-\Delta z_{c},z_{c}+\Delta z_{c}]$. 
This calculation yields: 
\begin{eqnarray}&&
\!\!\!\!\!\!
\Delta\Phi_{1}(t,z_{c})\!=\!-2\epsilon_{3}
\!\!\int_{z_{c}-\Delta z_{c}}^{z_{c}+\Delta z_{c}} 
\!\!\!\!\!\!\!\!\!\!\! dz' A_{1}(z') A_{2}^{2}(z')
\tilde\Psi_{10}(t,z')\tilde\Psi_{20}^{2}(t,z').
\nonumber \\&&
\label{coll5}
\end{eqnarray}  
The only function on the right hand side of Eq. (\ref{coll5}) 
that contains fast variations in $z$, which are of order 1, is $\tilde\Psi_{20}$. 
We can therefore approximate $A_{1}(z)$, $A_{2}(z)$, and $\tilde\Psi_{10}(t,z)$ 
by $A_{1}(z_{c}^{-})$, $A_{2}(z_{c}^{-})$, and $\tilde\Psi_{10}(t,z_{c})$, 
where $A_{j}(z_{c}^{-})$ is the limit from the left of $A_{j}$ at $z_{c}$. 
Furthermore, we can take into account only the fast dependence 
of $\tilde\Psi_{20}$ on $z$, i.e., the $z$ dependence that is contained in the factors  
$y=t-y_{20}-d_{1}z$. Denoting this approximation of 
$\tilde\Psi_{20}(t,z)$ by $\bar\Psi_{20}(y,z_{c})$, we obtain: 
\begin{eqnarray}&&
\!\!\!\!\!\!\!\!
\Delta\Phi_{1}(t,z_{c})\!=\!
-2\epsilon_{3}A_{1}(z_{c}^{-}) A_{2}^{2}(z_{c}^{-})
\tilde\Psi_{10}(t,z_{c})\times
\nonumber \\&&
\!\!\int_{z_{c}-\Delta z_{c}}^{z_{c}+\Delta z_{c}}  
\!\!\!\!\!\!\!\! dz' 
\bar\Psi_{20}^{2}(t-y_{20}-d_{1}z',z_{c}).
\label{coll5_add1}
\end{eqnarray}    
Since the integrand on the right hand side of Eq. (\ref{coll5_add1}) 
is sharply peaked at a small interval around $z_{c}$,  
we can extend the integral's limits to $-\infty$ and $\infty$. 
We also change the integration variable from 
$z'$ to $y=t-y_{20}-d_{1}z'$ and obtain: 
\begin{eqnarray} &&
% one-line version 1
\!\!\!\!\!\!\!
\Delta\Phi_{1}(t,z_{c})\!=\!-\frac{2\epsilon_{3}
A_{1}(z_{c}^{-}) A_{2}^{2}(z_{c}^{-})}{|d_{1}|}\tilde\Psi_{10}(t,z_{c})
\!\!\!\int_{-\infty}^{\infty} \!\!\!\!\! dy \bar\Psi_{20}^{2}(y,z_{c}).
\nonumber \\&&
\label{coll6}
\end{eqnarray}
In Appendix \ref{appendA}, we show that the net collision-induced amplitude shift 
of pulse 1 $\Delta A_{1}^{(c)}$ is related to the net collision-induced change 
in the envelope of the pulse  $\Delta\Phi_{1}(t,z_{c})$ by:
\begin{eqnarray}&&
\!\!\!\!\!\!\!\!\!\!\!\!\!\!
\Delta A_{1}^{(c)}=
\left[\int_{-\infty}^{\infty} \!\!\!\!\! dt \tilde\Psi_{10}^{2}(t,z_{c})\right]^{-1}
\!\!\int_{-\infty}^{\infty} \!\!\!\!\! dt \tilde\Psi_{10}(t,z_{c})\Delta\Phi_{1}(t,z_{c}). 
\label{coll6_add1}
\end{eqnarray}       
Substitution of Eq. (\ref{coll6}) into  Eq. (\ref{coll6_add1}) yields the following   
expression for the net collision-induced amplitude shift of pulse 1: 
\begin{eqnarray} &&
\!\!\!\!
\Delta A_{1}^{(c)}=-\frac{2\epsilon_{3}
A_{1}(z_{c}^{-}) A_{2}^{2}(z_{c}^{-})}{|d_{1}|}
\int_{-\infty}^{\infty} dy \bar\Psi_{20}^{2}(y,z_{c}).
\label{coll7}
\end{eqnarray}    
Note that 
\begin{eqnarray} &&
\!\!\!\!
\int_{-\infty}^{\infty} dy \bar\Psi_{20}^{2}(y,z_{c}) = 
\int_{-\infty}^{\infty} dt \tilde\Psi_{20}^{2}(t,z_{c}) = 
\int_{-\infty}^{\infty} dt |\tilde\psi_{20}(t,z_{c})|^{2}.
\nonumber
%\label{coll7_add2}
\end{eqnarray}    
But since $\int_{-\infty}^{\infty} dt |\tilde\psi_{20}(t,z)|^{2}$ is a conserved quantity 
of the unperturbed linear propagation equation, the following relations hold 
\begin{eqnarray} &&
\!\!\!\!
\int_{-\infty}^{\infty} dt |\tilde\psi_{20}(t,z_{c})|^{2} = 
\int_{-\infty}^{\infty} dt |\tilde\psi_{20}(t,0)|^{2} = 
\int_{-\infty}^{\infty} dt \tilde\Psi_{20}^{2}(t,0) = 
\mbox{const}.
\nonumber
%\label{coll7_add3}
\end{eqnarray}     
Thus, we can replace the integral on the right hand side of Eq. (\ref{coll7}) by 
$\int_{-\infty}^{\infty} dt \tilde\Psi_{20}^{2}(t,0)$ and obtain 
\begin{eqnarray} &&
\!\!\!\!
\Delta A_{1}^{(c)}=-\frac{2\epsilon_{3}
A_{1}(z_{c}^{-}) A_{2}^{2}(z_{c}^{-})}{|d_{1}|}
\int_{-\infty}^{\infty} dt \tilde\Psi_{20}^{2}(t,0).
% \int_{-\infty}^{\infty} dy \bar\Psi_{20}^{2}(y,0).
\label{coll7_add1}
\end{eqnarray}    
We note that the collision-induced amplitude shift $\Delta A_{1}^{(c)}$ 
depends only on the values of $A_{1}(z_{c}^{-})$, $A_{2}(z_{c}^{-})$, 
$|d_{1}|$, and on the initial total energy of pulse 2, $\int_{-\infty}^{\infty} dt \tilde\Psi_{20}^{2}(t,0)$.  
The amplitude shift does not depend on any other properties of the initial pulses. 
Thus, the expression for the amplitude shift is universal in the sense that it is 
independent of the exact details of the initial pulse shapes. 
Equation (\ref{coll7_add1}) is expected to hold for generic pulse shapes $\Psi_{j0}(t,z)$ 
with tails that decay sufficiently fast, such that the approximations leading from 
Eq. (\ref{coll5}) to Eq. (\ref{coll6}) are valid. Our numerical simulations with the 
coupled propagation model (\ref{coll1}), whose results are presented 
in Sec. \ref{waveguides_simu}, reveal that Eq. (\ref{coll7_add1}) 
is valid even for pulses with power-law decreasing tails, such as generalized Cauchy-Lorentz pulses, 
and for pulses that are initially nonsmooth and that develop significant tails during the propagation, 
such as square pulses.

We now use Eq. (\ref{coll7_add1}) to obtain expressions for the amplitude shifts   
in fast collisions between hyperbolic secant pulses, generalized Cauchy-Lorentz pulses, 
and square pulses, whose initial envelopes are given by Eqs. (\ref{IC1}), 
(\ref{IC2}), and (\ref{IC3}), respectively. We find that in all three cases, 
the amplitude shift of pulse 1 has the form 
\begin{eqnarray} &&
\Delta A_{1}^{(c)}=
-C_{P} \epsilon_{3} W_{20}
A_{1}(z_{c}^{-}) A_{2}^{2}(z_{c}^{-})/|d_{1}|, 
\label{coll8}
\end{eqnarray}             
where $W_{20}$ is the initial pulse width of pulse 2 and $C_{P}$ is a constant, 
whose value depends on the total initial energy of pulse 2.  
Furthermore, we find that $C_{P}=4$ for a collision between hyperbolic secant pulses, 
$C_{P}=3\pi/2^{7/4}$ for a collision between generalized Cauchy-Lorentz pulses, 
and $C_{P}=2$ for a collision between square pulses. 
In Ref. \cite{PNC2010}, we showed that the amplitude shift in a fast collision 
between two solitons of the NLS equation in the presence of weak cubic loss
is given by: $\Delta\eta_{1}^{(c)}=-4\epsilon_{3}\eta_{1}(z_{c}^{-})\eta_{2}(z_{c}^{-})/|\Delta\beta|$
(see Eq. (11) in Ref. \cite{PNC2010}).   
Noting that the soliton width is $W_{j}=1/\eta_{j}$, 
we can express the collision-induced amplitude shift of the soliton as:
\begin{eqnarray} &&
\Delta\eta_{1}^{(c)}=
-4\epsilon_{3}W_{2}(z_{c}^{-})\eta_{1}(z_{c}^{-})
\eta_{2}^{2}(z_{c}^{-})/|\Delta\beta| .
\label{coll8_add2}
\end{eqnarray}      
Thus, the expression for the collision-induced amplitude shift of the soliton  
has exactly the same form as the expression in Eq. (\ref{coll8}) for a collision 
between two pulses of the linear propagation equation. We also observe that 
$C_{P}=4$ in a two-soliton collision and in a collision between two hyperbolic 
secant pulses of the linear propagation equation.

\subsection{Numerical simulations for different pulse shapes}
\label{waveguides_simu} 

Since the prediction of section \ref{waveguides_delta_A} for universal behavior 
of the collision-induced amplitude shift in fast two-pulse collisions is based on several 
simplifying assumptions, it is important to check this prediction by numerical simulations 
with the full propagation model (\ref{coll1}). Equation (\ref{coll1}) is numerically integrated 
by employing the split-step method with periodic boundary conditions \cite{Agrawal2001,Yang2010}.    
For concreteness and without loss of generality, we present the results of  
the simulations with parameter values $\epsilon_{1}=0.01$, $\epsilon_{3}=0.01$, 
and $\mbox{sgn}(\tilde\beta_{2})=1$. 
Since we are interested in fast collisions, the values of $d_{1}$ are varied 
in the intervals $-60 \le d_{1} \le -2$ and $2 \le d_{1} \le 60$.  
To demonstrate the universal behavior of the collision-induced amplitude shift, 
we carry out the simulations with the three representative initial pulse shapes given 
by Eqs. (\ref{IC1})-(\ref{IC3}), which correspond to hyperbolic secant pulses, 
generalized Cauchy-Lorentz pulses, and square pulses.   
The values of the initial amplitudes, initial widths, initial phases, and 
initial position of pulse 1 are chosen as $A_{j}(0)=1$, $W_{j0}=4$, 
$\alpha_{j0}=0$, and $y_{10}=0$. 
The initial position of pulse 2 $y_{20}$ and the final propagation distance $z_{f}$ 
are chosen, such that the two pulses are well separated at $z=0$ and at $z=z_{f}$. 
In particular, we choose $y_{20}=\pm 25$ and $z_{f}=4$ for hyperbolic secant pulses, 
$y_{20}=\pm 20$ and $z_{f}=3$ for generalized Cauchy-Lorentz pulses, 
and $y_{20}=\pm 6$ and $z_{f}=4.5$  for square pulses. We emphasize, however, 
that results similar to the ones presented below are obtained in 
numerical simulations with other physical parameter values.      
For each of the three types of pulse shapes we present the dependence 
of $\Delta A_{1}^{(c)}$ on $d_{1}$ obtained in the simulations together 
with the analytic prediction of Eq. (\ref{coll8}). 
We also discuss the behavior of the relative error in the approximation of $\Delta A_{1}^{(c)}$, 
which is defined by $|\Delta A_{1}^{(c)(num)}-\Delta A_{1}^{(c)(th)}|\times 100/|\Delta A_{1}^{(c)(th)}|$.
The procedures used for obtaining the values of 
$\Delta A_{1}^{(c)}$ from Eq. (\ref{coll8}) and 
for calculating $\Delta A_{1}^{(c)}$ from the results of 
the numerical simulations are described in Appendix \ref{appendB}.

\begin{figure}[ptb]
\begin{tabular}{cc}
\epsfxsize=8.5cm  \epsffile{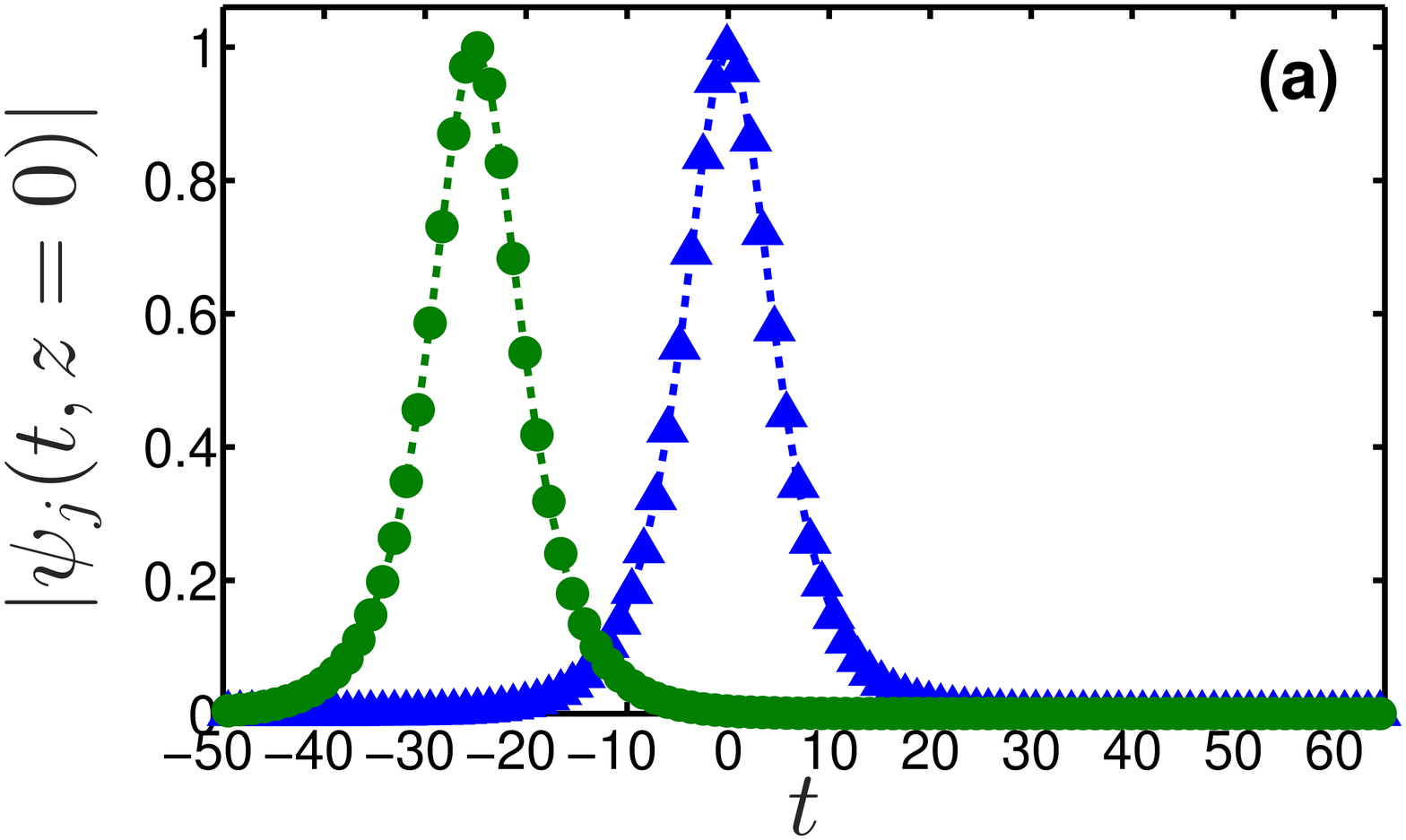} \\
\epsfxsize=8.5cm  \epsffile{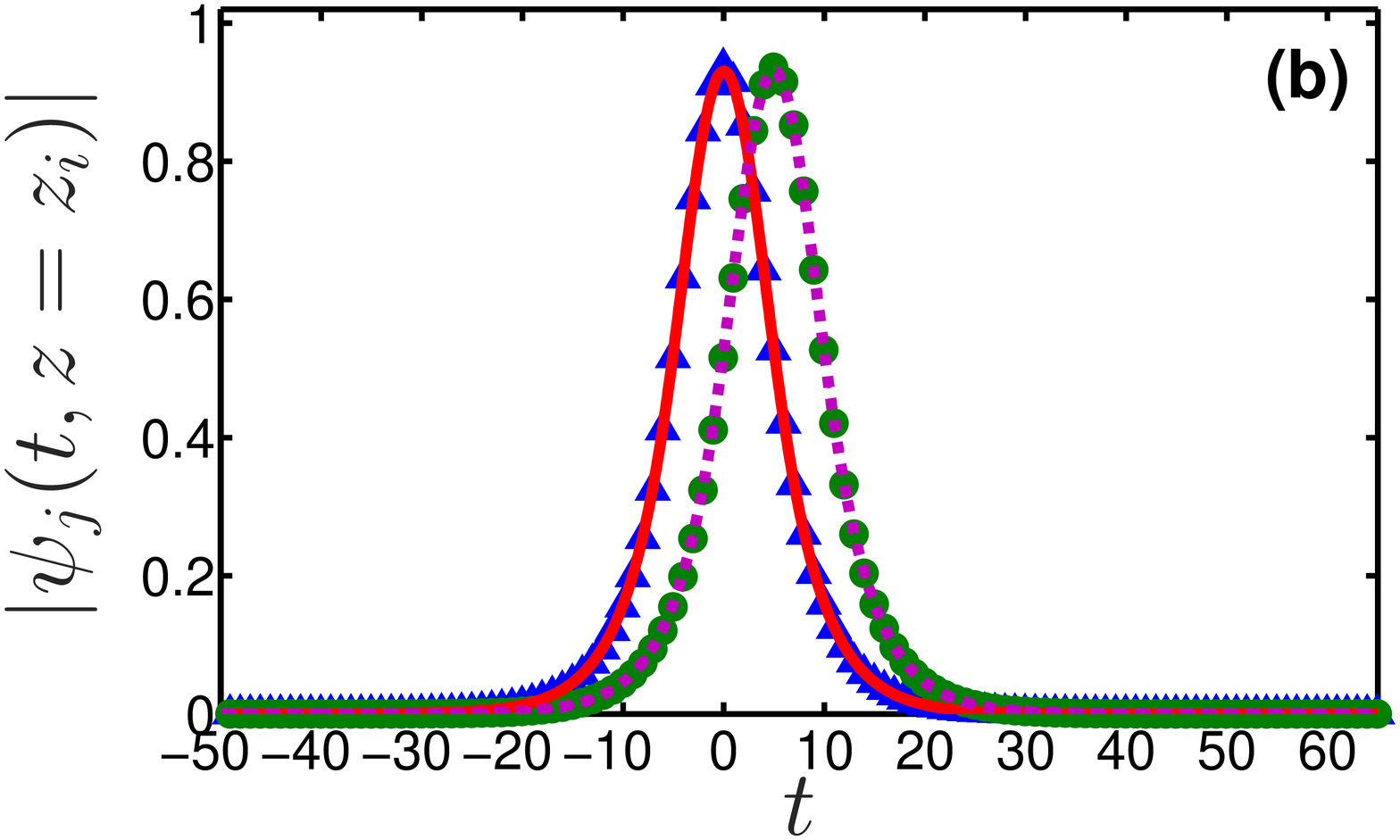} \\
\epsfxsize=8.5cm  \epsffile{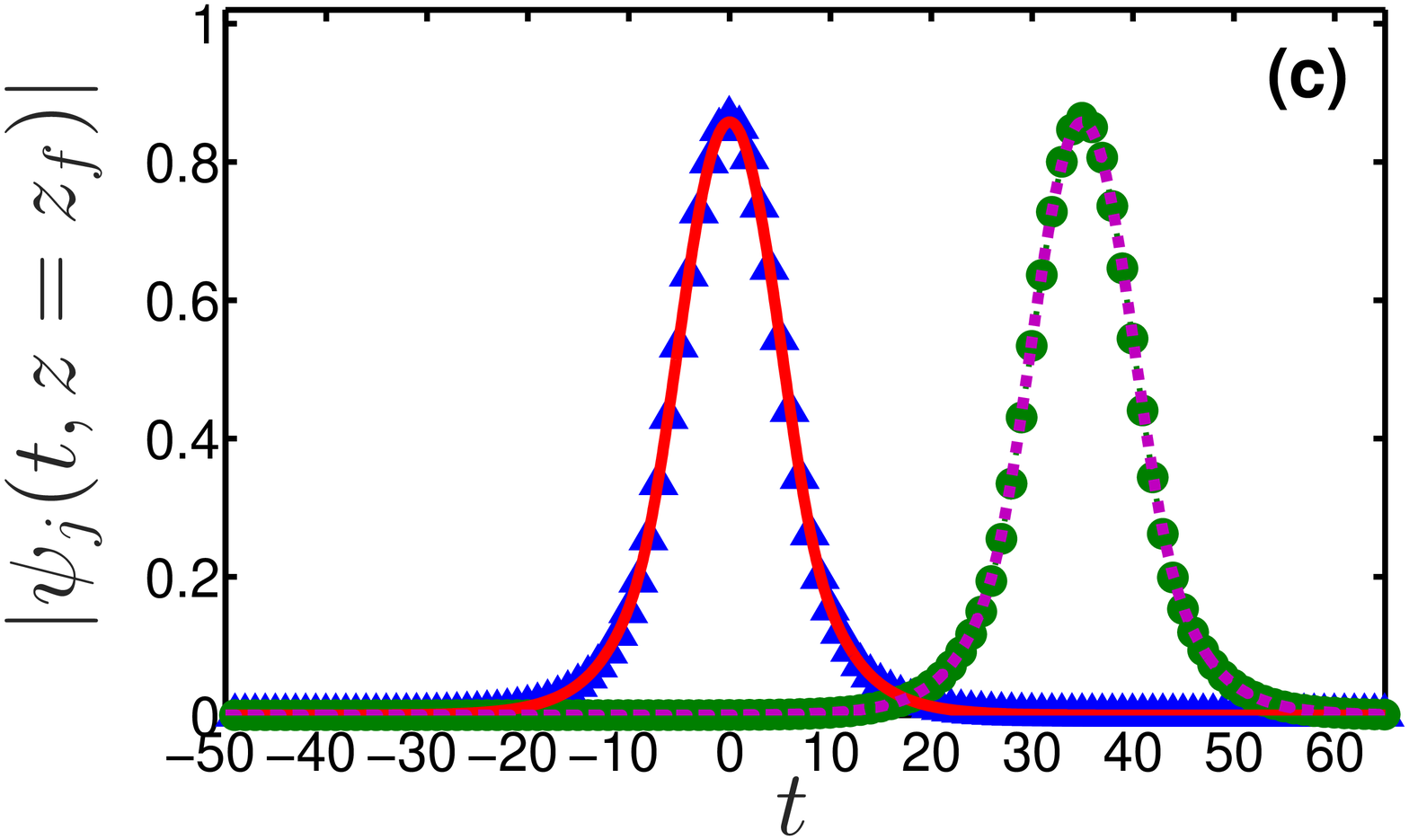} 
\end{tabular}
\caption{(Color online) The pulse shapes $|\psi_{j}(t,z)|$ at $z=0$ (a), $z=z_{i}=2$ (b), and $z=z_{f}=4$ (c) 
in a fast collision between two hyperbolic secant pulses in a linear waveguide 
with weak linear and cubic loss. The first-order dispersion coefficient is $d_{1}=15$. 
The blue triangles and green circles represent the initial pulse shapes $|\psi_{j}(t,0)|$ 
with $j=1,2$ in (a), and the perturbation theory's prediction for $|\psi_{j}(t,z)|$ with $j=1,2$ in (b) and (c). 
The solid red and dashed magenta curves in (b) and (c) correspond to $|\psi_{j}(t,z)|$ with $j=1,2$, 
obtained by numerical solution of Eq. (\ref{coll1}).}
\label{fig1}
\end{figure}

\begin{figure}[ptb]
\epsfxsize=12.0cm  \epsffile{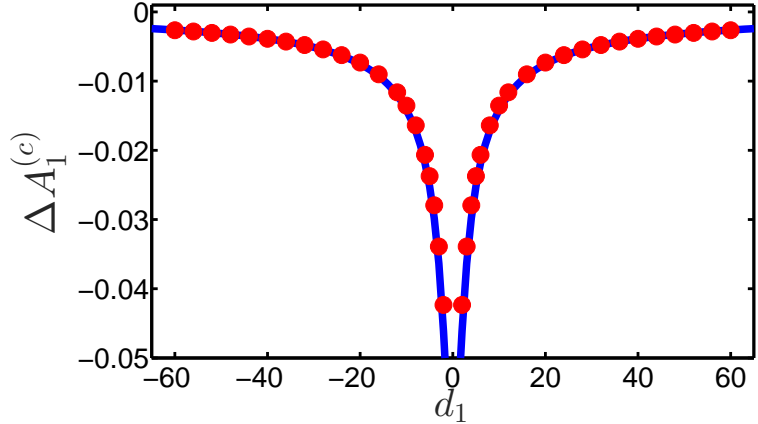} 
\caption{(Color online) The collision-induced amplitude shift of pulse 1 
$\Delta A_{1}^{(c)}$ vs the group velocity parameter $d_{1}$ 
in a fast collision between two hyperbolic secant pulses in a linear waveguide 
with weak linear and cubic loss. 
The red circles correspond to the result obtained by  
numerical solution of Eq. (\ref{coll1}). The solid blue curve represents 
the prediction of Eq. (\ref{coll8}) with $C_{P}=4$.}
 \label{fig2}
\end{figure}

We start by discussing the results of the numerical simulations for 
fast collisions between hyperbolic secant pulses. 
Figure \ref{fig1} shows the initial pulse shapes $|\psi_{j}(t,0)|$, 
and the pulse shapes $|\psi_{j}(t,z)|$ obtained in the simulation with $d_{1}=15$
at the intermediate distance $z_{i}=2>z_{c}$, and at the final distance $z_{f}=4$ \cite{zi_values}. 
Also shown is the analytic prediction for $|\psi_{j}(t,z)|$, which is obtained by employing Eq. (\ref{coll2}). 
We observe very good agreement between the analytic prediction 
and the result of the numerical simulation at both $z=z_{i}$ and $z=z_{f}$.   
In addition, we find that the pulses undergo broadening due to 
second-order dispersion and that no significant tail develops up to the final 
distance $z_{f}$. The dependence of the collision-induced amplitude shift 
$\Delta A_{1}^{(c)}$ on $d_{1}$ obtained by the numerical simulations 
is shown in Fig. \ref{fig2} along with the analytic prediction of Eq. (\ref{coll8}). 
The agreement between the analytic prediction and the simulations results is very good. 
More specifically, the relative error in the approximation of $\Delta A_{1}^{(c)}$ 
is less than 4.3$\%$ for $10 \le |d_{1}| \le 60$ 
and less than 8.2$\%$ for $2 \le |d_{1}| < 10$. Thus, the analytic prediction 
of Eq. (\ref{coll8}) provides a good approximation for the actual value of the 
collision-induced amplitude shift even at $d_{1}$ values that are not much larger than 1. 
These findings together with similar findings obtained in Ref. \cite{PNH2017B} for collisions 
between Gaussian pulses demonstrate the universal behavior of the amplitude 
shift in fast collisions between pulses with shapes that exhibit exponential or faster than 
exponential decrease with time.

\begin{figure}[ptb]
\begin{tabular}{cc}
\epsfxsize=8.5cm  \epsffile{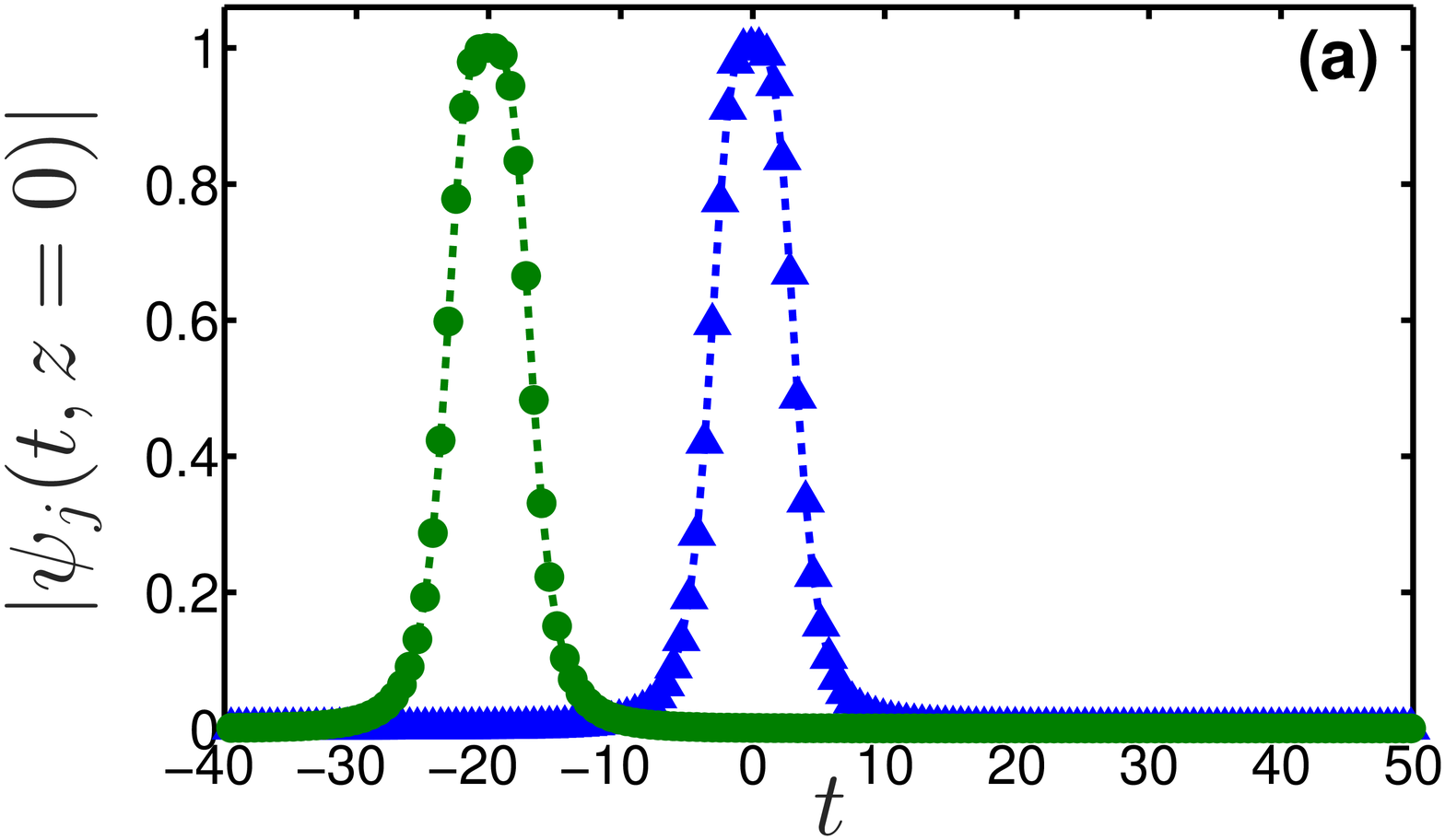} \\
\epsfxsize=8.5cm  \epsffile{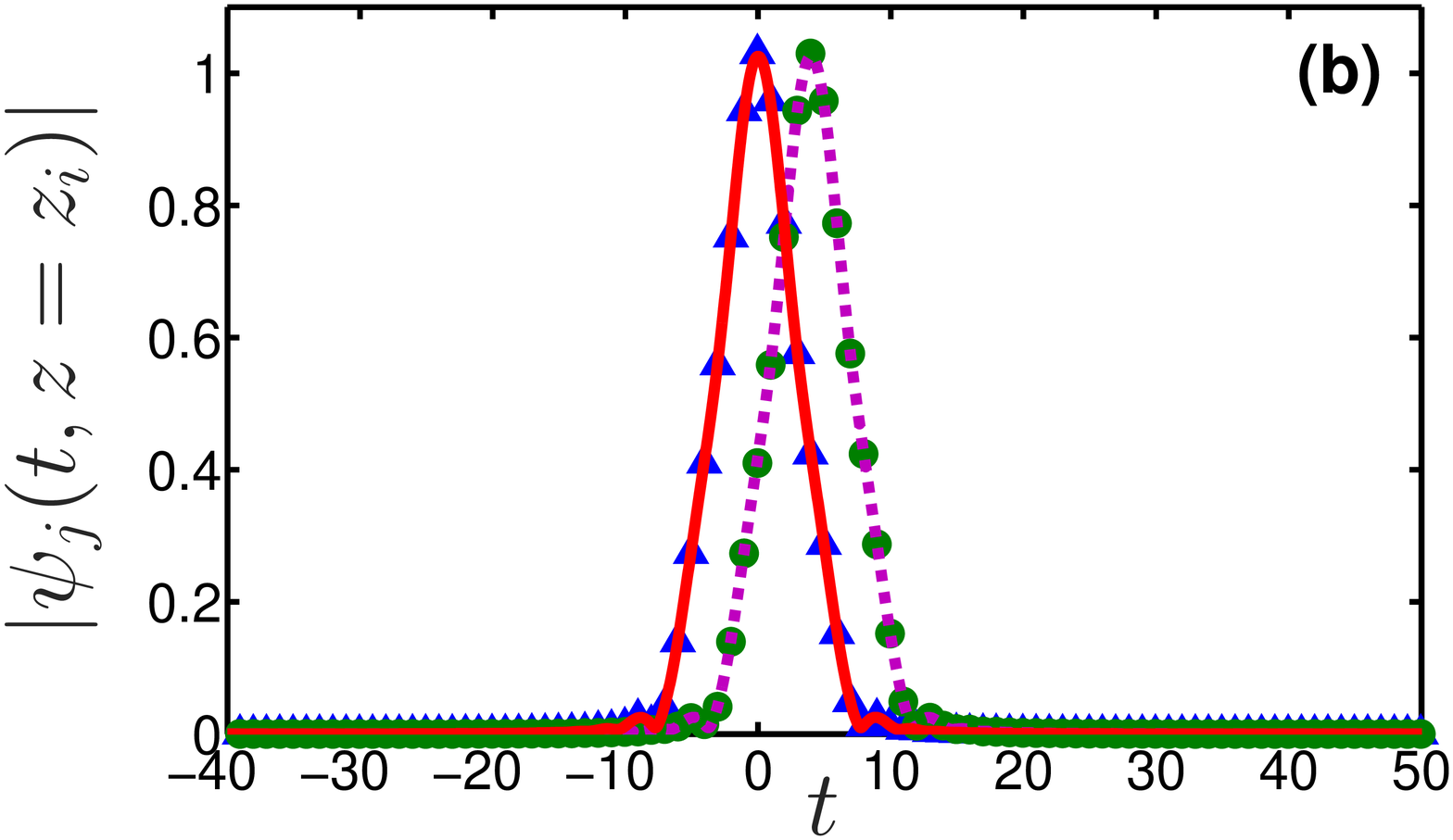} \\
\epsfxsize=8.5cm  \epsffile{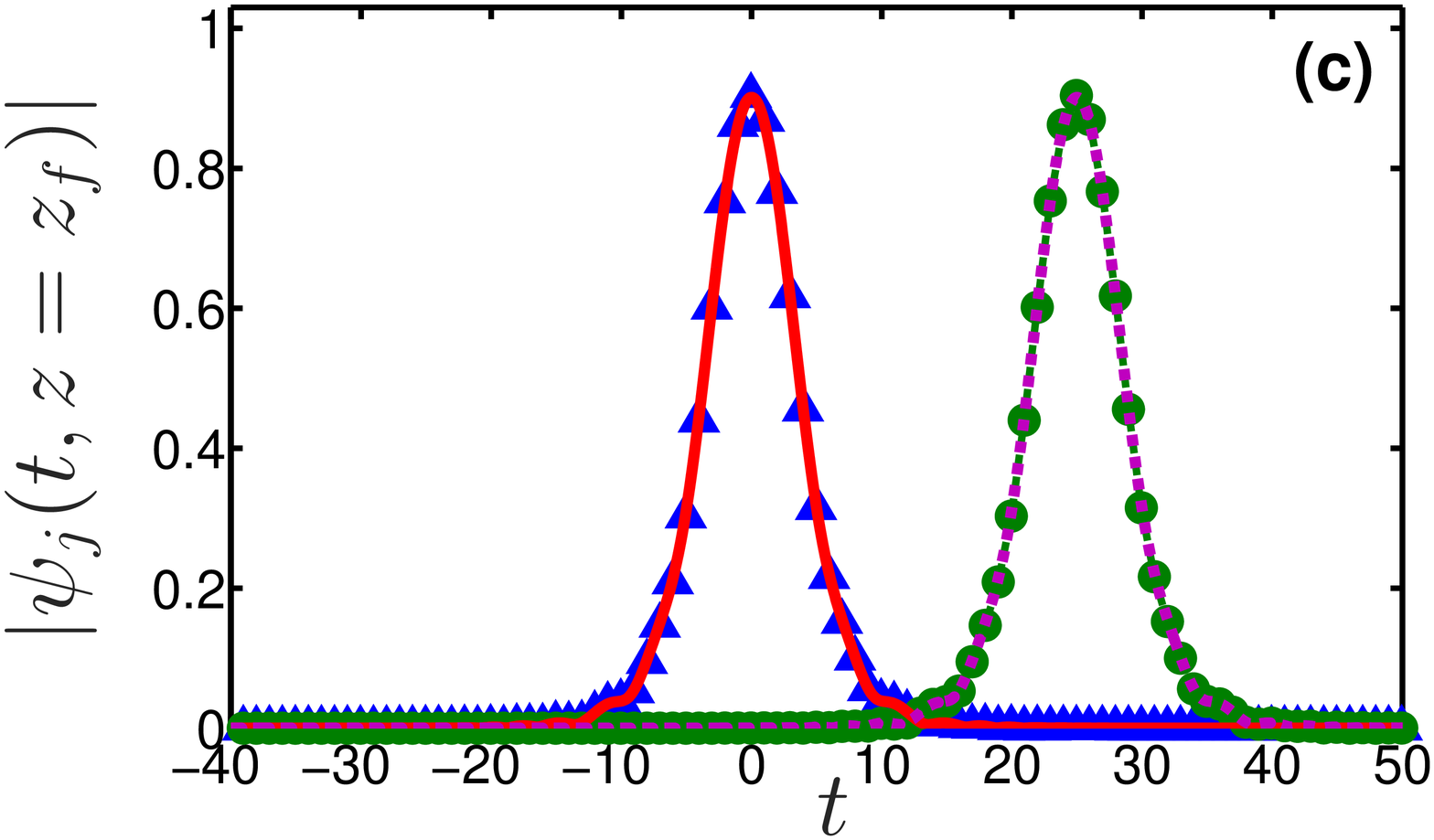} 
\end{tabular}
\caption{(Color online) The pulse shapes $|\psi_{j}(t,z)|$ at $z=0$ (a), $z=z_{i}=1.571$ (b), and $z=z_{f}=3$ (c) 
in a fast collision between two generalized Cauchy-Lorentz pulses in a linear waveguide 
with weak linear and cubic loss. The first-order dispersion coefficient is $d_{1}=15$. 
The blue triangles and green circles represent the initial pulse shapes $|\psi_{j}(t,0)|$ 
with $j=1,2$ in (a), and the perturbation theory's prediction for $|\psi_{j}(t,z)|$ with $j=1,2$ in (b) and (c). 
The solid red and dashed magenta curves in (b) and (c) correspond to $|\psi_{j}(t,z)|$ with $j=1,2$, 
obtained by numerical solution of Eq. (\ref{coll1}).}
 \label{fig3}
\end{figure}

\begin{figure}[ptb]
\epsfxsize=12.0cm  \epsffile{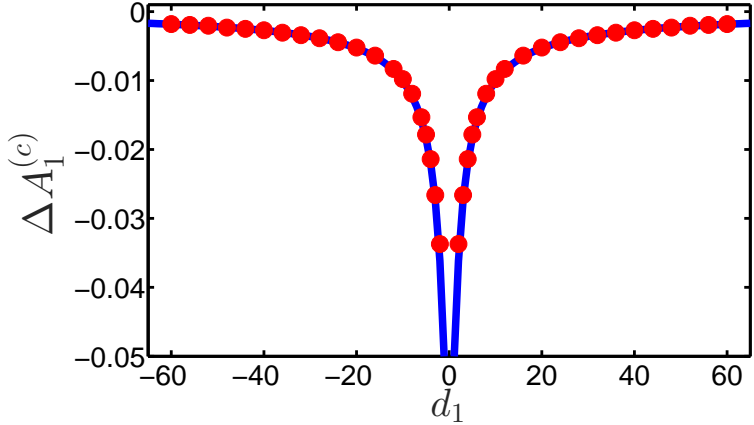} 
\caption{(Color online) The collision-induced amplitude shift of pulse 1 
$\Delta A_{1}^{(c)}$ vs the group velocity parameter $d_{1}$ 
in a fast collision between two generalized Cauchy-Lorentz pulses 
in a linear waveguide with weak linear and cubic loss. 
The red circles represent the result obtained by  
numerical solution of Eq. (\ref{coll1}). The solid blue curve represents 
the prediction of Eq. (\ref{coll8}) with $C_{P}=3\pi/2^{7/4}$.}
 \label{fig4}
\end{figure}

Next, we consider fast collisions between generalized Cauchy-Lorentz pulses. 
The initial pulse shapes $|\psi_{j}(t,0)|$, and the pulse shapes $|\psi_{j}(t,z)|$ 
obtained in the simulation with $d_{1}=15$ at $z_{i}=1.571>z_{c}$ and at $z_{f}=3$ 
are shown in Fig. \ref{fig3}. The analytic prediction for $|\psi_{j}(t,z)|$, 
which is obtained with Eq. (\ref{coll2}), is also shown. 
We observe that the pulses undergo considerable broadening 
and develop observable tails due to the effects of second-order dispersion. 
Despite of this, the agreement between the prediction of the perturbation theory 
and the simulation's result is very good at both $z=z_{i}$ and $z=z_{f}$.  
The dependence of $\Delta A_{1}^{(c)}$ on $d_{1}$ obtained in the simulations 
is shown in Fig. \ref{fig4} together with the analytic prediction of Eq. (\ref{coll8}). 
We observe very good agreement between the results of the simulations and the analytic 
prediction. In particular, the relative error in the approximation of $\Delta A_{1}^{(c)}$ 
is smaller than 2.9$\%$ for $10 \le |d_{1}| \le 60$ and smaller than 
6.8$\%$ for $2 \le |d_{1}| < 10$. Similar results are obtained for other values of the 
physical parameters and for other pulse shapes with power-law decreasing tails.                                                                            
Thus, our current study extends the results of Ref. \cite{PNH2017B}, in which it was assumed 
that the initial pulse shapes must possess tails that exhibit exponential or faster than 
exponential decrease with time for the perturbation theory to hold. 
Moreover, our results demonstrate that the universal behavior of the collision-induced 
amplitude shift is also observed in collisions between pulses with a relatively slow 
decay of the tails, such as power-law decay.

\begin{figure}[ptb]
\begin{tabular}{cc}
\epsfxsize=8.5cm  \epsffile{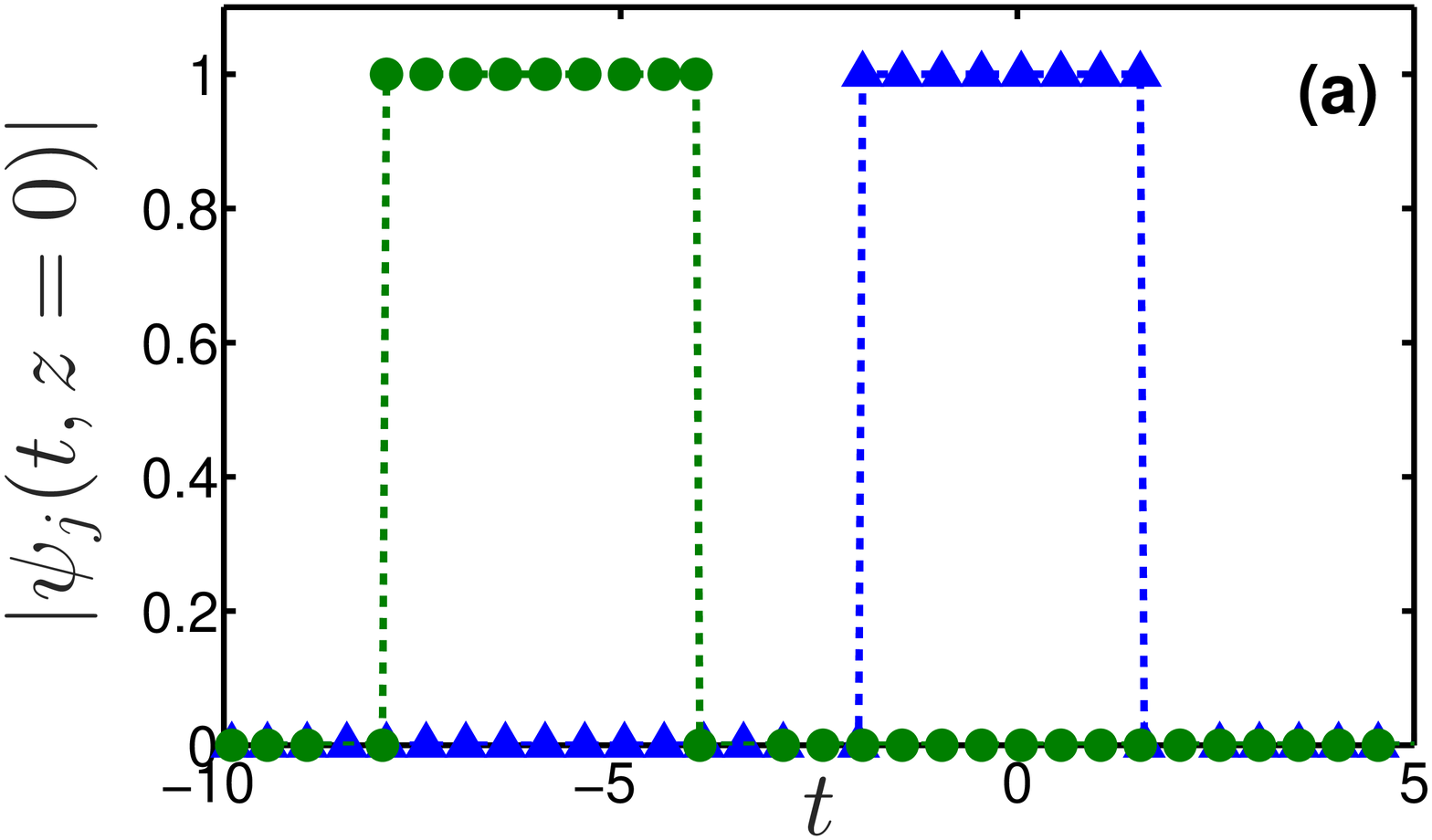} \\
\epsfxsize=8.5cm  \epsffile{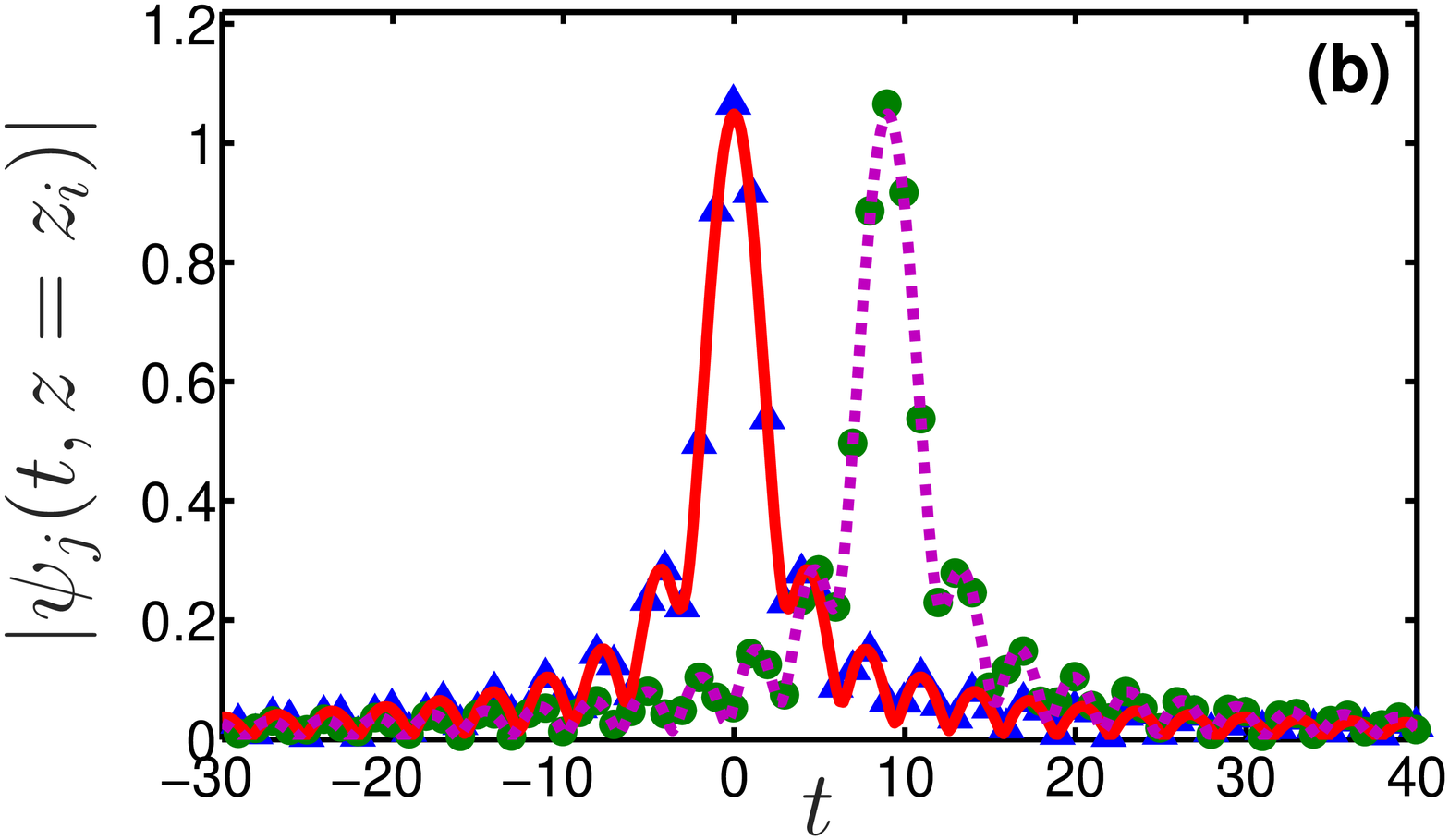} \\
\epsfxsize=8.5cm  \epsffile{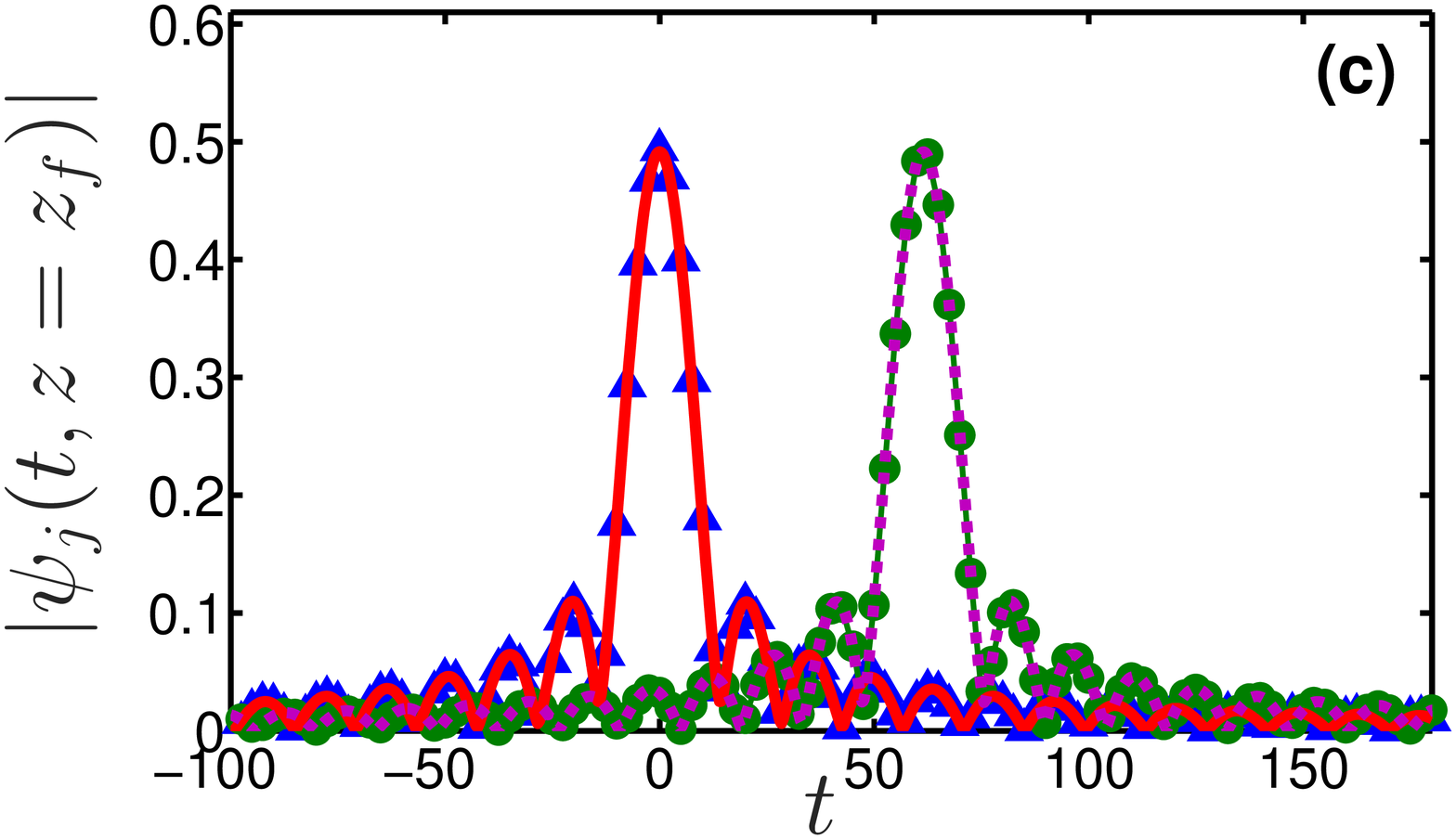} 
\end{tabular}
\caption{(Color online) The pulse shapes $|\psi_{j}(t,z)|$ at $z=0$ (a), $z=z_{i}=0.986$ (b), and $z=z_{f}=4.5$ (c) 
in a fast collision between two square pulses in a linear waveguide 
with weak linear and cubic loss. The first-order dispersion coefficient is $d_{1}=15$. 
The blue triangles and green circles represent the initial pulse shapes $|\psi_{j}(t,0)|$ 
with $j=1,2$ in (a), and the perturbation theory's prediction for $|\psi_{j}(t,z)|$ with $j=1,2$ in (b) and (c). 
The solid red and dashed magenta curves in (b) and (c) correspond to $|\psi_{j}(t,z)|$ with $j=1,2$, 
obtained by numerical solution of Eq. (\ref{coll1}).}
 \label{fig5}
\end{figure}

\begin{figure}[ptb]
\epsfxsize=12.0cm  \epsffile{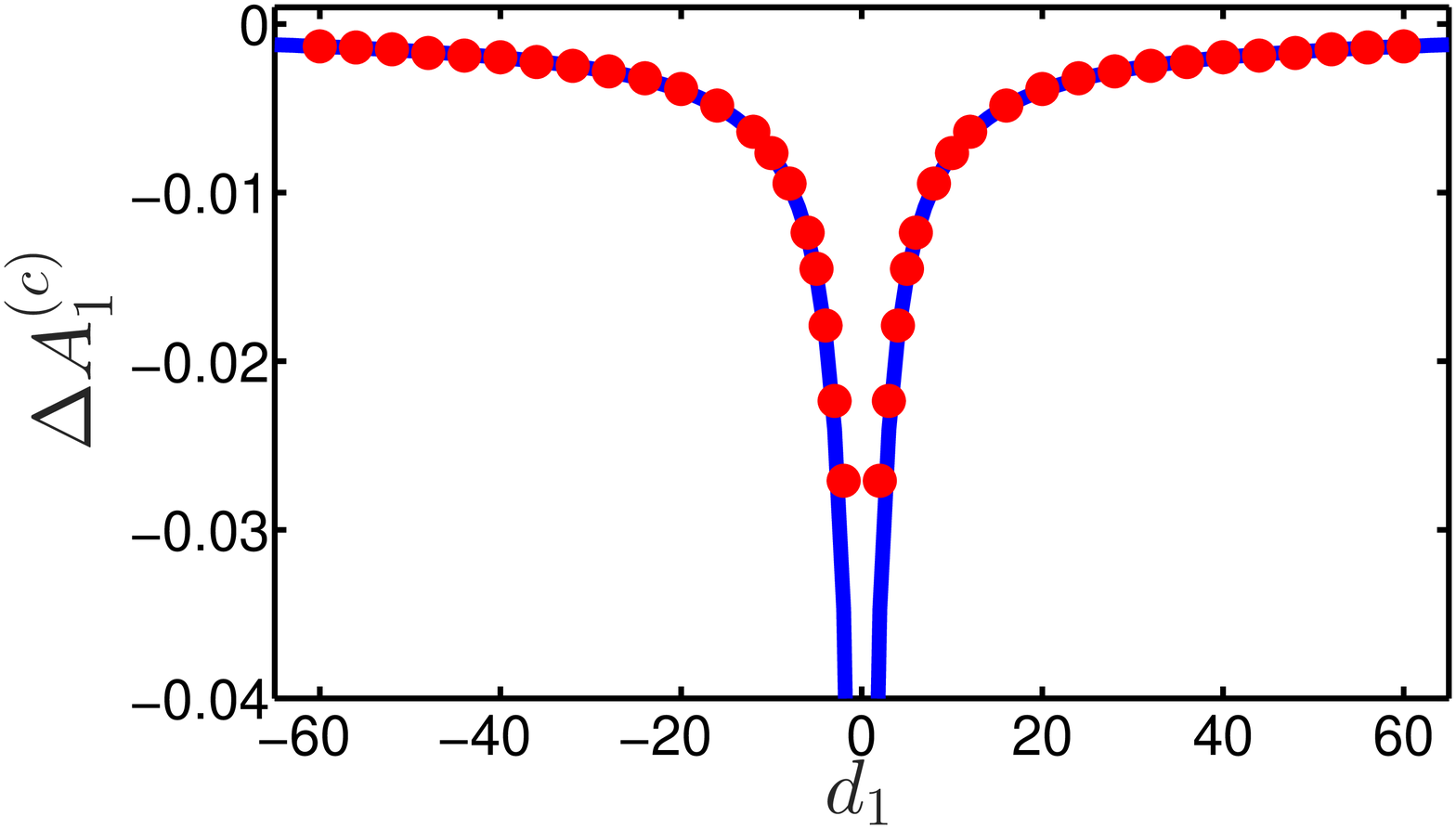} 
\caption{(Color online) The collision-induced amplitude shift of pulse 1 
$\Delta A_{1}^{(c)}$ vs the group velocity parameter $d_{1}$ 
in a fast collision between two square pulses 
in a linear waveguide with weak linear and cubic loss. 
The red circles represent the result obtained by  
numerical simulations with Eq. (\ref{coll1}). The solid blue curve corresponds to 
the prediction of Eq. (\ref{coll8}) with $C_{P}=2$.}
 \label{fig6}
\end{figure}

We now turn to consider fast collisions between square pulses. 
The initial pulse shapes $|\psi_{j}(t,0)|$, and the pulse shapes $|\psi_{j}(t,z)|$ 
obtained in the simulation with $d_{1}=15$ at the intermediate distance  
$z_{i}=0.986>z_{c}$ and at the final distance $z_{f}=4.5$ 
are shown in Fig. \ref{fig5}. The analytic prediction obtained with Eq. (\ref{coll2}) is also shown. 
We observe that the pulses develop extended oscillatory tails and experience broadening 
due to the effects of second-order dispersion. The development of the extended tails 
occurs on a length scale in $z$, which is smaller than the collision length $\Delta z_{c}$. 
Despite of this fact, we observe very good agreement between the perturbation theory's 
prediction for the pulse shapes and the simulation's result at both $z=z_{i}$ and $z=z_{f}$. 
The dependence of the collision-induced amplitude shift 
$\Delta A_{1}^{(c)}$ on $d_{1}$ obtained in the simulations 
is shown in Fig. \ref{fig6} along with the analytic prediction of Eq. (\ref{coll8}). 
The agreement between the analytic prediction and the results of the simulations is very good, 
despite of the development of extended pulse tails. 
More specifically, the relative error in the approximation of $\Delta A_{1}^{(c)}$ 
is smaller than 3.8$\%$ for $10 \le |d_{1}| \le 60$, smaller than 7.2$\%$ for $3 \le |d_{1}| < 10$, 
and is equal to 22.0$\%$ at $|d_{1}|=2$. The good agreement between the analytic prediction 
and the results of the numerical simulations can be explained in the following manner.  
First, since $\Delta z_{c} \ll z_{D}$, most of the energy is contained in the main bodies 
of the pulses during the collision [see Fig. \ref{fig5}(b)]. 
Second, the total energy integrals $\int_{-\infty}^{\infty} dt \tilde\Psi_{j0}^{2}(t,z)$, 
which appear in the calculation of $\Delta A_{1}^{(c)}$ [see Eqs. (\ref{coll6_add1})-(\ref{coll7_add1})], 
are conserved by the unperturbed linear propagation equation. As a result, the redistribution of the total energy of the 
pulses due to the development of extended tails does not have a significant effect on the values of 
$\Delta A_{1}^{(c)}$ measured in the simulations for $|d_{1}| \ge 3$, 
as long as the main bodies of the pulses are well-separated at $z_{f}$. 
Instead, contributions to the amplitude shift coming from interaction between the main body of one pulse 
and the tail of the other pulse and between the tails of both pulses partially compensate for the reduction 
in the contribution coming from direct interaction between the main bodies of the two pulses. 
Based on the results presented in Figs. \ref{fig5} and \ref{fig6} and on similar results obtained 
with other values of the physical parameters, we conclude that the universal behavior of 
the collision-induced amplitude shift can be observed even in collisions between pulses, 
which develop extended tails during the collision.

\section{Fast collisions in systems described by coupled linear 
diffusion-advection models}
\label{diffusion}
\subsection{Evolution model and initial pulse shapes}
\label{diffusion_model}
We consider the dynamics of a fast collision between pulses 
of two substances, denoted by 1 and 2, that evolve in the presence of 
linear diffusion, weak linear and quadratic loss, and advection of 
material 2 with velocity $v_{d}$ relative to material 1.  
The dynamics of the fast two-pulse collision is described by   
the following system of perturbed coupled linear diffusion-advection equations \cite{PNH2017B}: 
\begin{eqnarray} &&
\!\!\!\!\!\!\!\!\!\!\!\!\!\!
\partial _{t}u_{1}=\partial _{x}^{2}u_{1}-\varepsilon_{1}u_{1}
-\varepsilon_{2}u_{1}^{2}-2\varepsilon_{2}u_{1}u_{2},
\nonumber \\&&
\!\!\!\!\!\!\!\!\!\!\!\!\!\!
\partial _{t}u_{2}=\partial _{x}^{2}u_{2}-v_{d}\partial _{x}u_{2}
-\varepsilon_{1}u_{2}-\varepsilon_{2}u_{2}^{2}-2\varepsilon_{2}u_{1}u_{2},
\label{rda1}
\end{eqnarray}         
where $u_{1}$ and $u_{2}$ are the concentrations of substance 1 and 2,  
$t$ is time, $x$ is a spatial coordinate, and the linear and quadratic loss coefficients
$\varepsilon_{1}$ and $\varepsilon_{2}$ satisfy $0<\varepsilon_{1} \ll 1$ 
and $0<\varepsilon_{2} \ll 1$ \cite{Dimensions2}.     
The term $-v_{d}\partial _{x}u_{2}$ in Eq. (\ref{rda1}) describes advection, 
while the terms $-\varepsilon_{1}u_{j}$ describe the effects of linear loss. 
The terms $-\varepsilon_{2}u_{j}^{2}$ and $-2\varepsilon_{2}u_{j}u_{k}$ 
describe intra-substance and inter-substance effects due to quadratic loss, respectively.  
Note that in Eq. (\ref{rda1}) we assume that the nonlinear loss is weak, and therefore, 
the effects of higher-order loss can be neglected. We point out that the effects of higher-order loss 
on the collision-induced amplitude shift can be calculated in a manner 
similar to the one described in Sec. \ref{diffusion_delta_A} 
(see also, Ref. \cite{QMN2018}, where a similar calculation was performed 
for collisions between Gaussian pulses in weakly perturbed linear optical waveguides).

We are interested in demonstrating universal behavior of the collision-induced amplitude 
shift. For this purpose, we consider fast collisions between pulses with generic 
initial shapes and with tails that decay sufficiently fast, 
such that the values of the integrals $\int_{-\infty}^{\infty} dx u_{j}(x,0)$ 
are finite. We assume that the pulses can be characterized by 
initial amplitudes $A_{j}(0)$, initial widths $W_{j0}$, and initial positions $x_{j0}$. 
Similar to Sec. \ref{waveguides}, we demonstrate the universal behavior 
of the collision-induced amplitude shift by considering the following three major types of pulses: 
(1) pulses with exponentially decreasing tails, (2) pulses with power-law decreasing 
tails, (3) pulses that are initially nonsmooth. For concreteness, we demonstrate the behavior 
of the amplitude shift using the following representative initial pulses: 
hyperbolic secant pulses in (1), generalized Cauchy-Lorentz pulses in (2), 
and square pulses in (3). The initial concentrations $u_{j}(x,0)$ for these 
pulses are given by  
\begin{eqnarray} &&
u_{j}(x,0)=A_{j}(0)\sech\left[(x-x_{j0})/W_{j0}\right],
\label{rda_IC1}
\end{eqnarray}   
with $j=1,2$ for hyperbolic secant pulses, by 
\begin{eqnarray} &&
u_{j}(x,0)=\frac{A_{j}(0)}
{1+ 2 \left[(x-x_{j0})/W_{j0} \right]^{4}} ,
\label{rda_IC2}
\end{eqnarray}
with $j=1,2$ for generalized Cauchy-Lorentz pulses, and by 
\begin{eqnarray} &&
\!\!\!\!\!\!\!\!\!
u_{j}(x,0)=
\left\{\begin{array}{l l}
A_{j}(0) & \;\; \mbox{for} \;\;  |x-x_{j0}| \le W_{j0}/2,\\
0 & \;\; \mbox{for} \;\; |x-x_{j0}| > W_{j0}/2,\\
\end{array} \right. 
\label{rda_IC3}
\end{eqnarray}     
with $j=1,2$ for square pulses. We point out that similar behavior of the 
collision-induced amplitude shift is observed for other choices of the initial 
pulse shapes.

\subsection{Calculation of the collision-induced amplitude shift}  
\label{diffusion_delta_A}
To obtain the expressions for the collision-induced amplitude shifts, 
we assume a complete fast two-pulse collision. 
The complete collision assumption means that the two pulses 
are well separated at $t=0$ and at the final time $t=t_{f}$.  
The assumption of a fast collision means that the collision time interval 
$\Delta t_{c}=W_{0}/|v_{d}|$, which is the time interval during which the two pulses overlap, 
is much shorter than the diffusion time $t_{D}=W_{0}^{2}$.   
Requiring $\Delta t_{c} \ll t_{D}$, we obtain $W_{0}|v_{d}| \gg 1$, 
as the condition for a fast collision.

We now demonstrate that the condition $W_{0}|v_{d}| \gg 1$ 
for a fast collision can be realized in weakly perturbed physical systems 
described by linear diffusion-advection models. An important example for these 
systems is provided by binary gas mixtures. For concreteness, consider diffusion of $\mbox{O}_{2}$ 
in $\mbox{N}_{2}$  at 293.15$^{\circ}$K and at a pressure of 1 atm. 
The diffusion coefficient is $D=0.202$ $\mbox{cm}^{2}/\mbox{s}$ \cite{CRC2004}.  
Thus, for an initial pulse width of 1 cm and for an advection velocity value of $V_{d}=3$ $\mbox{cm}/\mbox{s}$,
we find that $W_{0}|v_{d}| =14.85$. Similar results are obtained in other binary gas mixtures. 
For example, the diffusion coefficient of $\mbox{CO}_{2}$ in $\mbox{N}_{2}$ at 293.15$^{\circ}$K 
and at a pressure of 1 atm is $D=0.160$ $\mbox{cm}^{2}/\mbox{s}$ \cite{CRC2004}. 
Using this value we find that for an initial pulse width of 1 cm and for $V_{d}=3$ $\mbox{cm}/\mbox{s}$, 
$W_{0}|v_{d}| =18.75$. Therefore, the condition $W_{0}|v_{d}| \gg 1$ for a fast two-pulse 
collision is satisfied in both cases.

The perturbation technique for calculating the collision-induced amplitude shift is 
similar to the one derived in section \ref{waveguides_delta_A} 
for treating fast collisions between pulses of the linear propagation equation.   
Thus, we look for a solution of Eq. (\ref{rda1}) in the form 
\begin{eqnarray}&&
\!\!\!\!\!\!\!
u_{j}(x,t)=u_{j0}(x,t)+\phi_{j}(x,t), 
\label{rda2}
\end{eqnarray}      
where $j=1,2$, $u_{j0}$ are solutions of Eq. (\ref{rda1}) 
without inter-pulse interaction, and $\phi_{j}$ describe collision-induced effects. 
By definition, $u_{10}$ and $u_{20}$ satisfy the following two weakly perturbed 
linear diffusion equations  
\begin{eqnarray}&&
\!\!\!\!\!\!\!
\partial _{t}u_{10}=\partial _{x}^{2}u_{10}-\varepsilon_{1}u_{10}
-\varepsilon_{2}u_{10}^{2},
\!\!\!\!\!\!\!\!
\label{rda2_add1}
\end{eqnarray}         
and 
\begin{eqnarray}&&
\!\!\!\!\!\!\!
\partial _{t}u_{20}=\partial _{x}^{2}u_{20}-v_{d}\partial _{x}u_{20}
-\varepsilon_{1}u_{20}-\varepsilon_{2}u_{20}^{2}.
\!\!\!\!\!\!\!\!
\label{rda2_add2}
\end{eqnarray}  
We substitute the ansatz (\ref{rda2}) into Eq. (\ref{rda1}) 
and use Eqs. (\ref{rda2_add1}) and  (\ref{rda2_add2}) 
to obtain equations for $\phi_{1}$ and $\phi_{2}$.
We concentrate on the calculation of $\phi_{1}$, since the calculation of $\phi_{2}$ is similar.     
Taking into account only leading-order effects of the collision, 
we can neglect terms that contain products of $\varepsilon_{1}$ or $\varepsilon_{2}$ 
with $\phi_{1}$ or $\phi_{2}$, such as $-\varepsilon_{1}\phi_{1}$, 
$-2\varepsilon_{2}u_{10}\phi_{1}$, $-2\varepsilon_{2}u_{20}\phi_{1}$, 
$-2\varepsilon_{2}u_{10}\phi_{2}$, etc. We therefore obtain: 
\begin{equation}
\partial_{t}\phi_{1}=\partial _{x}^{2}\phi_{1}-2\varepsilon_{2}u_{10}u_{20}. 
\label{rda3} 
\end{equation}                                    
The term $-2\varepsilon_{2}u_{10}u_{20}$ on the right hand side 
of Eq. (\ref{rda3}) is of order $\varepsilon_{2}$.  
Additionally, the collision time interval $\Delta t_{c}$ is of order $1/|v_{d}|$  
and therefore, the term $\partial_{t}\phi_{1}$ is of order $|v_{d}| \times O(\phi_{1})$. 
Equating the orders of $\partial_{t}\phi_{1}$ and $-2\varepsilon_{2}u_{10}u_{20}$, 
we find that $\phi_{1}$ is of order $\varepsilon_{2}/|v_{d}|$.  
As a result, the term  $\partial _{x}^{2}\phi_{1}$ is of order 
$\varepsilon_{2}/|v_{d}|$ and can be neglected.   
Thus, the equation for the collision-induced change of pulse 1 
in the leading order of the perturbative calculation is:    
\begin{equation}
\partial _{t}\phi_{1}=-2\varepsilon_{2}u_{10}u_{20}.
\label{rda4} 
\end{equation}                         
Equation  (\ref{rda4}) is similar to Eq. (\ref{coll4}) in section \ref{waveguides_delta_A}   
and also to the equation obtained in Ref. \cite{PNC2010} for a fast collision between 
two solitons of the NLS equation in the presence of weak cubic loss.

The net collision-induced amplitude shift of pulse 1 is calculated from  
the net collision-induced change in the concentration of pulse 1. 
We denote by $t_{c}$ the collision time, 
which is the time at which the maxima of $u_{j}(x,t)$ coincide.  
In a fast collision, the collision takes place in the small 
time interval $[t_{c}-\Delta t_{c},t_{c}+\Delta t_{c}]$ around $t_{c}$.  
Therefore, the net collision-induced change in the concentration of pulse 1 
$\Delta\phi_{1}(x,t_{c})$ can be estimated by: 
$\Delta\phi_{1}(x,t_{c})=\phi_{1}(x,t_{c}+\Delta t_{c})-
\phi_{1}(x,t_{c}-\Delta t_{c})$. To calculate $\Delta\phi_{1}(x,t_{c})$,    
we introduce the approximation $u_{j0}(x,t)=A_{j}(t) \tilde u_{j0}(x,t)$, 
where $\tilde u_{j0}(x,t)$ is the solution 
of the unperturbed linear diffusion equation with unit amplitude.                                                                                                                                           
Substituting the approximate expressions for $u_{j0}$ into Eq. (\ref{rda4}) 
and integrating with respect to time over the interval 
$[t_{c}-\Delta t_{c},t_{c}+\Delta t_{c}]$, we obtain: 
\begin{eqnarray} &&
\!\!\!\!\!\!\!\!\!\!\!
\Delta\phi_{1}(x,t_{c})\!=\!
-2\varepsilon_{2}\!\!\int_{t_{c}-\Delta t_{c}}^{t_{c}+\Delta t_{c}} 
\!\!\!\!\!\!\!\!\!\! dt' A_{1}(t') A_{2}(t')
\tilde u_{10}(x,t') \tilde u_{20}(x,t'). 
\nonumber \\&&
\label{rda5}
\end{eqnarray}                                                                                 
The only function on the right hand side of Eq. (\ref{rda5}) 
that contains fast variations in $t$, which are of order 1, is $\tilde u_{20}$. 
Therefore, we can approximate $A_{1}(t)$, $A_{2}(t)$, 
and $\tilde u_{10}(x,t)$ by $A_{1}(t_{c}^{-})$, $A_{2}(t_{c}^{-})$, 
and $\tilde u_{10}(x,t_{c})$. Additionally, we can  
take into account only the fast dependence of 
$\tilde u_{20}$ on $t$, i.e., the $t$ dependence that is contained 
in the factors $y=x-x_{20}-v_{d}t$. Denoting this approximation of 
$\tilde u_{20}(x,t)$ by $\bar u_{20}(y,t_{c})$ 
and implementing the approximations, we obtain:
\begin{eqnarray} &&
\!\!\!\!\!\!\!\!\!\!\!
\Delta\phi_{1}(x,t_{c})\!=\!
-2\varepsilon_{2}A_{1}(t_{c}^{-})A_{2}(t_{c}^{-})\tilde u_{10}(x,t_{c}) \times
\nonumber \\&&
\!\!\!\int_{t_{c}-\Delta t_{c}}^{t_{c}+\Delta t_{c}} 
\!\!\!\!\!\!\!\! dt'  \bar u_{20}(x-x_{20}-v_{d}t',t_{c}). 
\label{rda5_add1}
\end{eqnarray}  
Since the integrand on the right hand side of Eq. (\ref{rda5_add1}) 
is sharply peaked at a small interval about $t_{c}$, we can extend 
the integral's limits to $-\infty$ and $\infty$. 
In addition, we change the integration variable from $t'$ to 
$y=x-x_{20}-v_{d}t'$ and obtain
\begin{eqnarray} &&
\!\!\!\!\!\! 
\Delta\phi_{1}(x,t_c)=
-\frac{2\varepsilon_{2}A_{1}(t_{c}^{-})A_{2}(t_{c}^{-})}{|v_{d}|}\tilde u_{10}(x,t_{c}) 
\!\!\!\int_{-\infty}^{\infty} \!\!\!\!\!\!\!\! dy \, \bar u_{20}(y,t_{c}).
\nonumber \\&&
%\;\;\;\;\;\;\;\;\;\;\;\;\;\;
%\times
\label{rda6}
\end{eqnarray}    
In Appendix \ref{appendA}, we show that the net collision-induced 
amplitude shift of pulse 1 $\Delta A_{1}^{(c)}$ is related to 
$\Delta\phi_{1}(x,t_c)$ by:
\begin{eqnarray}&&
\!\!\!\!\!\!\!\!\!\!\!\!\!\!
\Delta A_{1}^{(c)}=\left[\int_{-\infty}^{\infty} \!\!\!\!\! dx \, 
\tilde u_{10}(x,t_{c})\right]^{-1}
\!\!\int_{-\infty}^{\infty} \!\!\!\!\! dx \, 
\Delta\phi_{1}(x,t_c).
\label{rda6_add1}
\end{eqnarray}       
Substituting Eq. (\ref{rda6}) into  Eq. (\ref{rda6_add1}),  we arrive at 
the following expression for the total collision-induced amplitude shift of pulse 1: 
\begin{eqnarray} &&
\!\!\!\!\!\!\!\!
\Delta A_{1}^{(c)}=-\frac{2\varepsilon_{2}A_{1}(t_{c}^{-})A_{2}(t_{c}^{-})}{|v_{d}|}
\int_{-\infty}^{\infty} dy \, \bar u_{20}(y,t_{c}).
\label{rda7}
\end{eqnarray}    
We note that 
\begin{eqnarray} &&
\!\!\!\!
\int_{-\infty}^{\infty} dy \, \bar u_{20}(y,t_{c}) = 
\int_{-\infty}^{\infty} dx \, \tilde u_{20}(x,t_{c}).
\nonumber
\end{eqnarray}    
In addition, since $\int_{-\infty}^{\infty} dx \, \tilde u_{20}(x,t)$ 
is a conserved quantity of the unperturbed linear diffusion equation, 
the following relations hold 
\begin{eqnarray} &&
\!\!\!\!
\int_{-\infty}^{\infty} dx \, \tilde u_{20}(x,t_{c}) = 
\int_{-\infty}^{\infty} dx \, \tilde u_{20}(x,0) = 
\mbox{const}.
\nonumber
\end{eqnarray}     
Therefore, we can replace the integral on the right hand side of Eq. (\ref{rda7}) by 
$\int_{-\infty}^{\infty} dx \, \tilde u_{20}(x,0)$ and obtain 
\begin{eqnarray} &&
\!\!\!\!
\Delta A_{1}^{(c)}=-\frac{2\varepsilon_{2}A_{1}(t_{c}^{-})A_{2}(t_{c}^{-})}{|v_{d}|}
\int_{-\infty}^{\infty} dx \, \tilde u_{20}(x,0). 
\label{rda7_add1}
\end{eqnarray}    
We observe that the amplitude shift $\Delta A_{1}^{(c)}$ 
depends only on the values of $A_{1}(t_{c}^{-})$, $A_{2}(t_{c}^{-})$, 
$|v_{d}|$, and on the initial total mass of pulse 2, $\int_{-\infty}^{\infty} dx \, \tilde u_{20}(x,0)$.  
Since the expression for $\Delta A_{1}^{(c)}$ is independent of the exact details of the initial pulse shapes, 
we say that this expression is universal. 
Equation (\ref{rda7_add1}) is expected to hold for generic pulse shapes $u_{j0}(x,t)$ 
with tails that decay sufficiently fast, such that the approximations leading from 
Eq. (\ref{rda5}) to Eq. (\ref{rda6}) are valid.   
In Sec. \ref{diffusion_simu}, we show by numerical simulations with the 
coupled diffusion-advection model (\ref{rda1}) that Eq. (\ref{rda7_add1}) 
is valid even for pulses with power-law decreasing tails, such as generalized Cauchy-Lorentz pulses, 
and for pulses that are initially nonsmooth, such as square pulses.

Using Eq. (\ref{rda7_add1}), we can obtain explicit expressions for the amplitude shifts   
in fast collisions between hyperbolic secant pulses, generalized Cauchy-Lorentz pulses, 
and square pulses, whose initial shapes are given by Eqs. (\ref{rda_IC1}), 
(\ref{rda_IC2}), and (\ref{rda_IC3}), respectively. 
We find that in all three cases, $\Delta A_{1}^{(c)}$ is given by: 
\begin{eqnarray}
\!\!\!\!\!\!\!\!
\Delta A_{1}^{(c)}=-C_{D} \varepsilon_{2} W_{20}
A_{1}(t_{c}^{-}) A_{2}(t_{c}^{-})/|v_{d}|,
\label{rda8}
\end{eqnarray}                   
where the value of the constant $C_{D}$ depends on the initial total mass of pulse 2.  
In addition, we find that $C_{D}=2\pi$ for a collision between hyperbolic secant pulses, 
$C_{D}=2^{1/4}\pi$ for a collision between generalized Cauchy-Lorentz pulses, 
and $C_{D}=2$ for a collision between square pulses. 
We point out that Eqs. (\ref{rda7_add1}) and (\ref{rda8}) are similar to 
Eqs. (\ref{coll7_add1}) and (\ref{coll8}) for the amplitude shift in a fast collision  
between two pulses of the linear propagation model in the presence of weak cubic loss. 
Equation (\ref{rda8}) is also similar to Eq. (\ref{coll8_add2}) for the amplitude shift in a fast collision 
between two solitons of the NLS equation in the presence of  weak cubic loss.

\subsection{Numerical simulations for different pulse shapes}
\label{diffusion_simu}
To validate the predictions for universal behavior of the amplitude shift 
in fast two-pulse collisions, we carry out numerical simulations with Eq. (\ref{rda1}). 
The equation is numerically solved by the split-step method 
with periodic boundary conditions \cite{Verwer2003}. 
For concreteness and without loss of generality, we present here the results of simulations 
with parameter values $\varepsilon_{1}=0.01$ and $\varepsilon_{2}=0.01$. 
Since we are interested in fast collisions, the values of $v_{d}$ are varied 
in the intervals $-60 \le v_{d} \le -2$ and $2 \le v_{d} \le 60$.  
The universal behavior of the collision-induced amplitude shift is demonstrated 
by carrying out numerical simulations with the three typical initial pulse shapes given 
by Eqs. (\ref{rda_IC1})-(\ref{rda_IC3}), i.e., with hyperbolic secant pulses, 
generalized Cauchy-Lorentz pulses, and square pulses.   
The values of the initial amplitudes, initial widths, and initial position of pulse 1 are chosen 
as $A_{j}(0)=1$, $W_{j0}=4$, and $x_{10}=0$. 
The initial position of pulse 2 $x_{20}$ and the final time $t_{f}$ are chosen, 
such that the pulses are well separated at $t=0$ and at $t=t_{f}$. 
More specifically, we choose $x_{20}=\pm 25$ and $t_{f}=4$ for hyperbolic secant pulses, 
$x_{20}=\pm 25$ and $t_{f}=3.5$ for generalized Cauchy-Lorentz pulses, 
and $x_{20}=\pm 6$ and $t_{f}=1.5$  for square pulses. 
We point out  that results similar to the ones presented below are obtained in 
numerical simulations with other physical parameter values.      
For each pulse shape type, we present the dependence 
of $\Delta A_{1}^{(c)}$ on $v_{d}$ obtained in the simulations along 
with the perturbation theory's prediction of Eq. (\ref{rda8}). 
We also discuss the behavior of the relative error in the approximation of $\Delta A_{1}^{(c)}$. 
The procedures for calculating the values of 
$\Delta A_{1}^{(c)}$ from Eq. (\ref{rda8}) and from the results of 
the numerical simulations are similar to the ones described in Appendix \ref{appendB}.

\begin{figure}[ptb]
\begin{tabular}{cc}
\epsfxsize=8.5cm  \epsffile{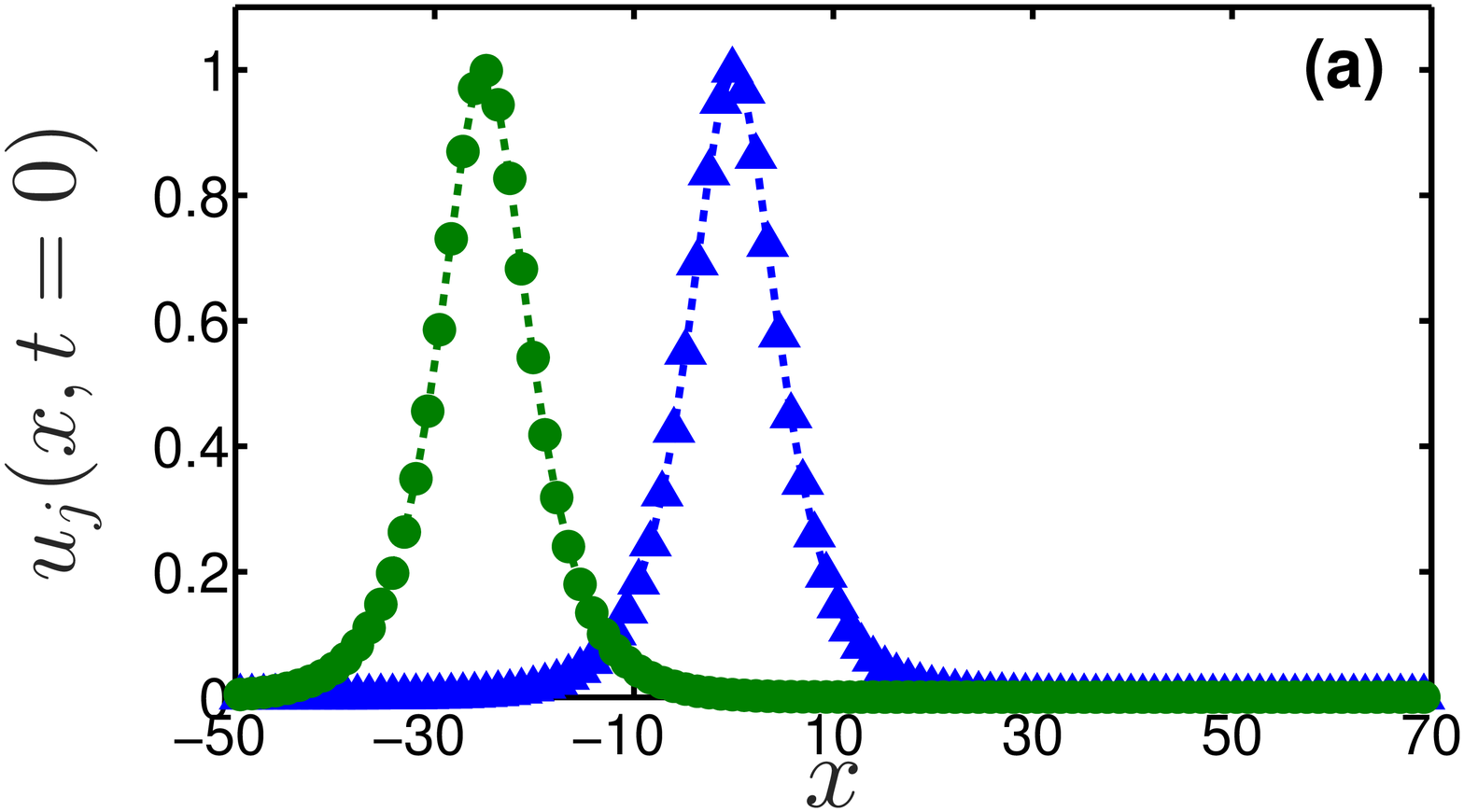} \\
\epsfxsize=8.5cm  \epsffile{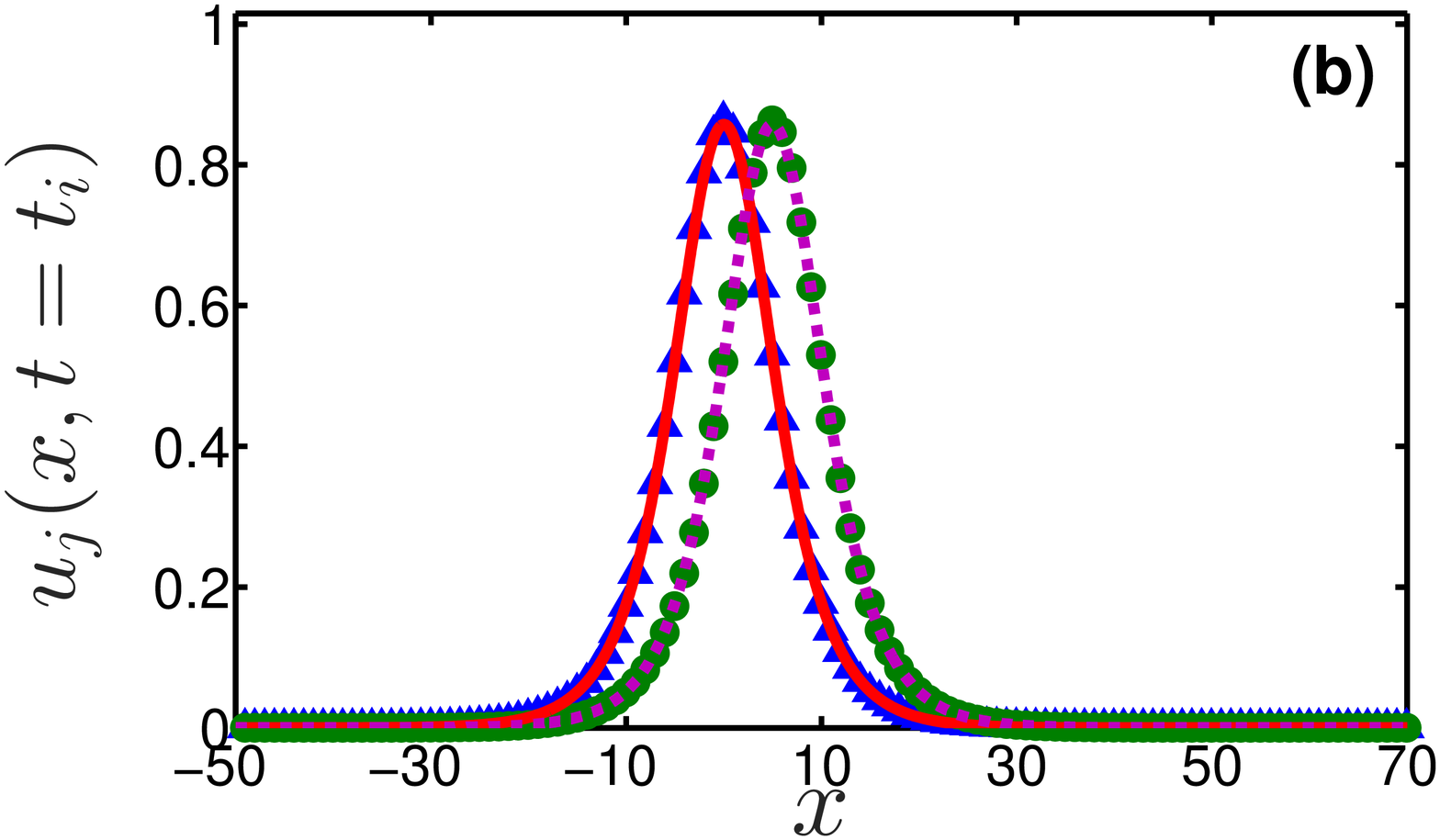} \\
\epsfxsize=8.5cm  \epsffile{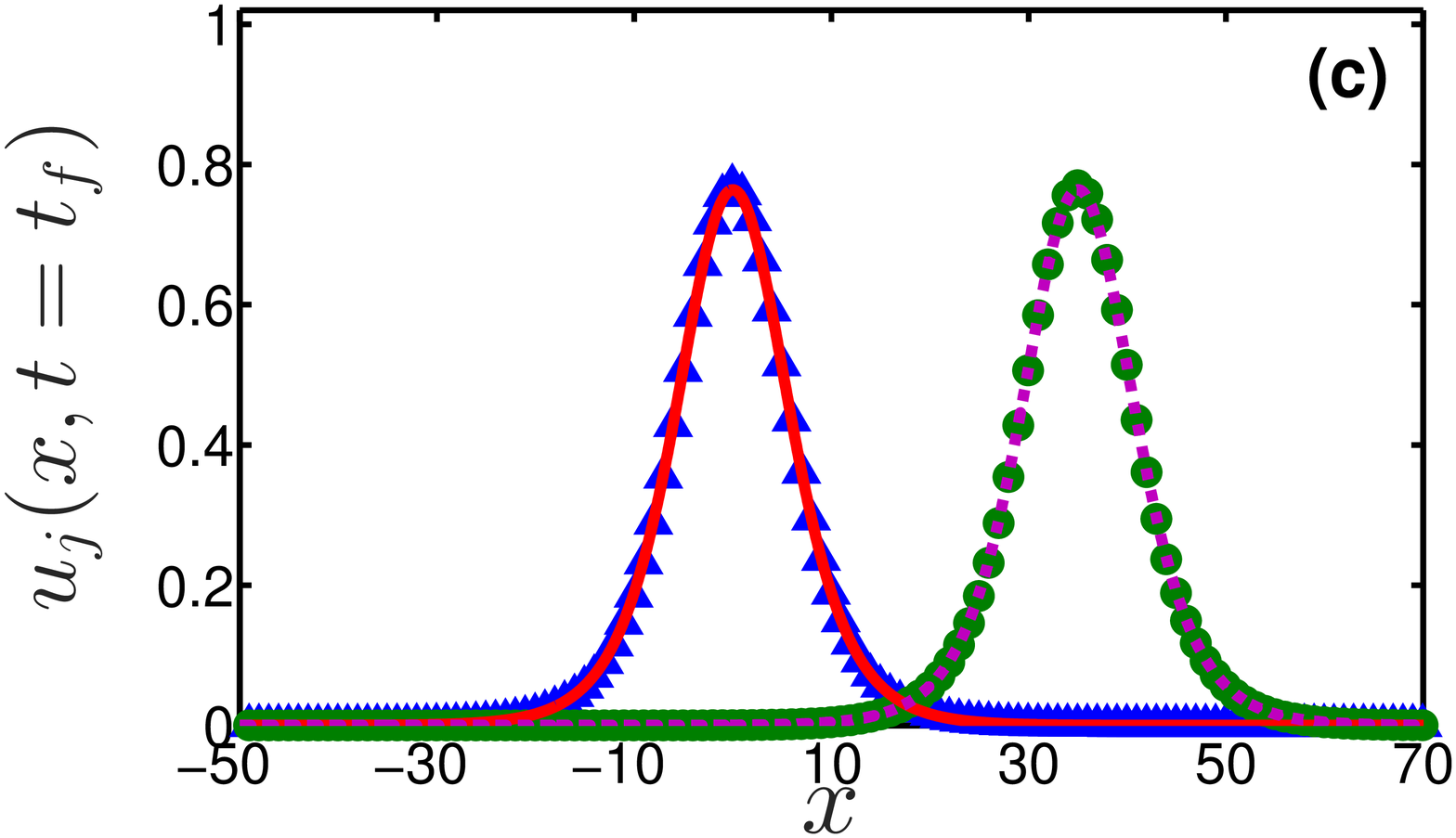} 
\end{tabular}
\caption{(Color online) The pulse shapes $u_j(x,t)$ at $t=0$ (a), $t=t_{i}=2$ (b), and $t=t_{f}=4$ (c) 
in a fast collision between two hyperbolic secant pulses in a system described by the 
coupled diffusion-advection model (\ref{rda1}). The advection velocity is $v_{d}=15$. 
The blue triangles and green circles represent the initial pulse shapes $u_j(x,0)$ with $j=1,2$ in (a), 
and the perturbation theory's prediction for $u_j(x,t)$ with $j=1,2$ in (b) and (c). 
The solid red and dashed magenta curves in (b) and (c) correspond to $u_j(x,t)$ with $j=1,2$, 
obtained by numerical solution of Eq. (\ref{rda1}).}
\label{fig7}
\end{figure}

\begin{figure}[ptb]
\epsfxsize=12.0cm  \epsffile{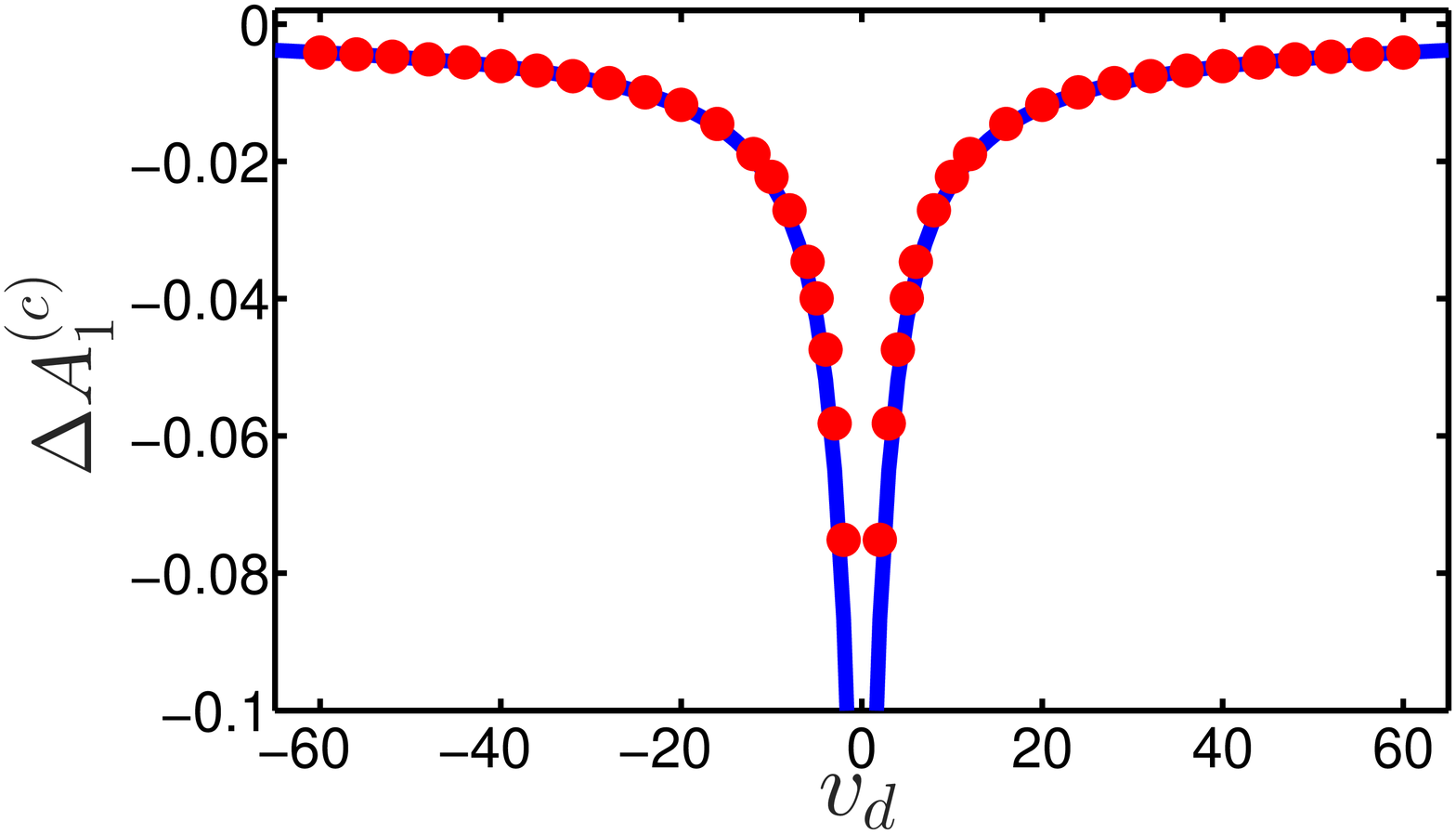} 
\caption{(Color online) The collision-induced amplitude shift of pulse 1 
$\Delta A_{1}^{(c)}$ vs advection velocity $v_{d}$ 
in a fast collision between two hyperbolic secant pulses in a system  
described by the coupled diffusion-advection model (\ref{rda1}).  
The red circles represent the result obtained by numerical solution of Eq. (\ref{rda1}). 
The solid blue curve corresponds to the prediction of Eq. (\ref{rda8}) with $C_{D}=2\pi$.}
\label{fig8}
\end{figure}

We first discuss the results of the numerical simulations for fast collisions 
between hyperbolic secant pulses. Figure \ref{fig7} shows the pulse shapes 
$u_j(x,t)$ obtained in the simulation with $v_{d}=15$ at $t=0$, 
at the intermediate time $t_{i}=2>t_{c}$, and at the final time $t_{f}=4$ \cite{ti_values}. 
The analytic prediction for $u_j(x,t)$, obtained with Eq. (\ref{rda2}), is also shown. 
We observe that the pulses experience broadening due to diffusion. 
Despite of the broadening, the agreement between the numerical result 
and the analytic prediction is very good at both $t=t_{i}$ and $t=t_{f}$.   
Figure \ref{fig8} shows the dependence of $\Delta A_{1}^{(c)}$ on $v_{d}$ 
obtained by the simulations along with the analytic prediction of Eq. (\ref{rda8}). 
The agreement between the analytic prediction and the simulations results is very good. 
In particular, the relative error in the approximation of $\Delta A_{1}^{(c)}$ 
is smaller than 4.1$\%$ for $10 \le |v_{d}| \le 60$ 
and smaller than 13.3$\%$ for $2 \le |v_{d}| < 10$.
Similar behavior was observed in Ref. \cite{PNH2017B} for collisions 
between Gaussian pulses. Thus, our findings in the current paper and in Ref. \cite{PNH2017B} 
demonstrate the universal behavior of the amplitude shift in fast collisions between pulses, 
whose tails exhibit exponential or faster than exponential decrease with increasing distance 
from the pulse maximum.

\begin{figure}[ptb]
\begin{tabular}{cc}
\epsfxsize=8.5cm  \epsffile{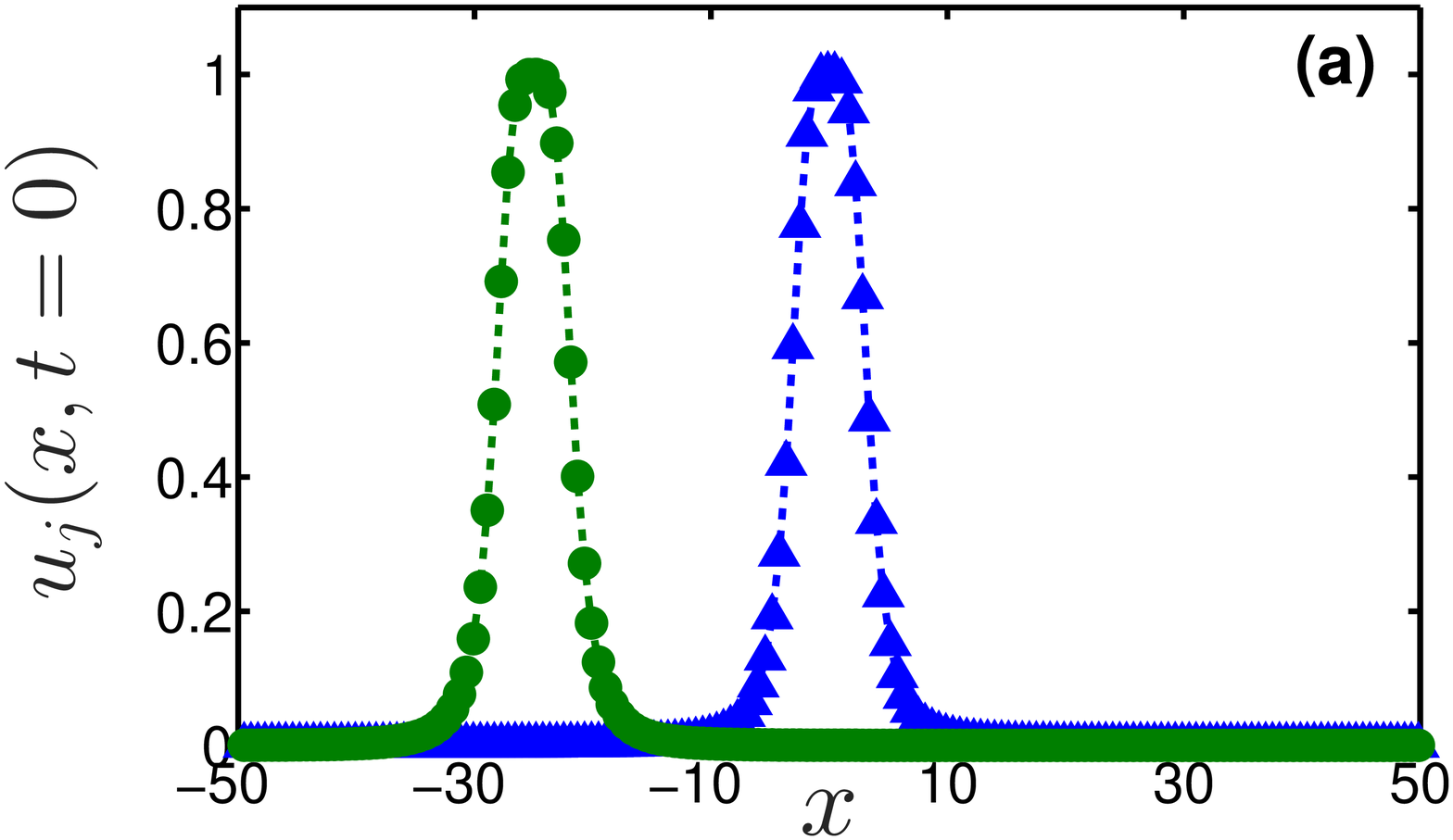} \\
\epsfxsize=8.5cm  \epsffile{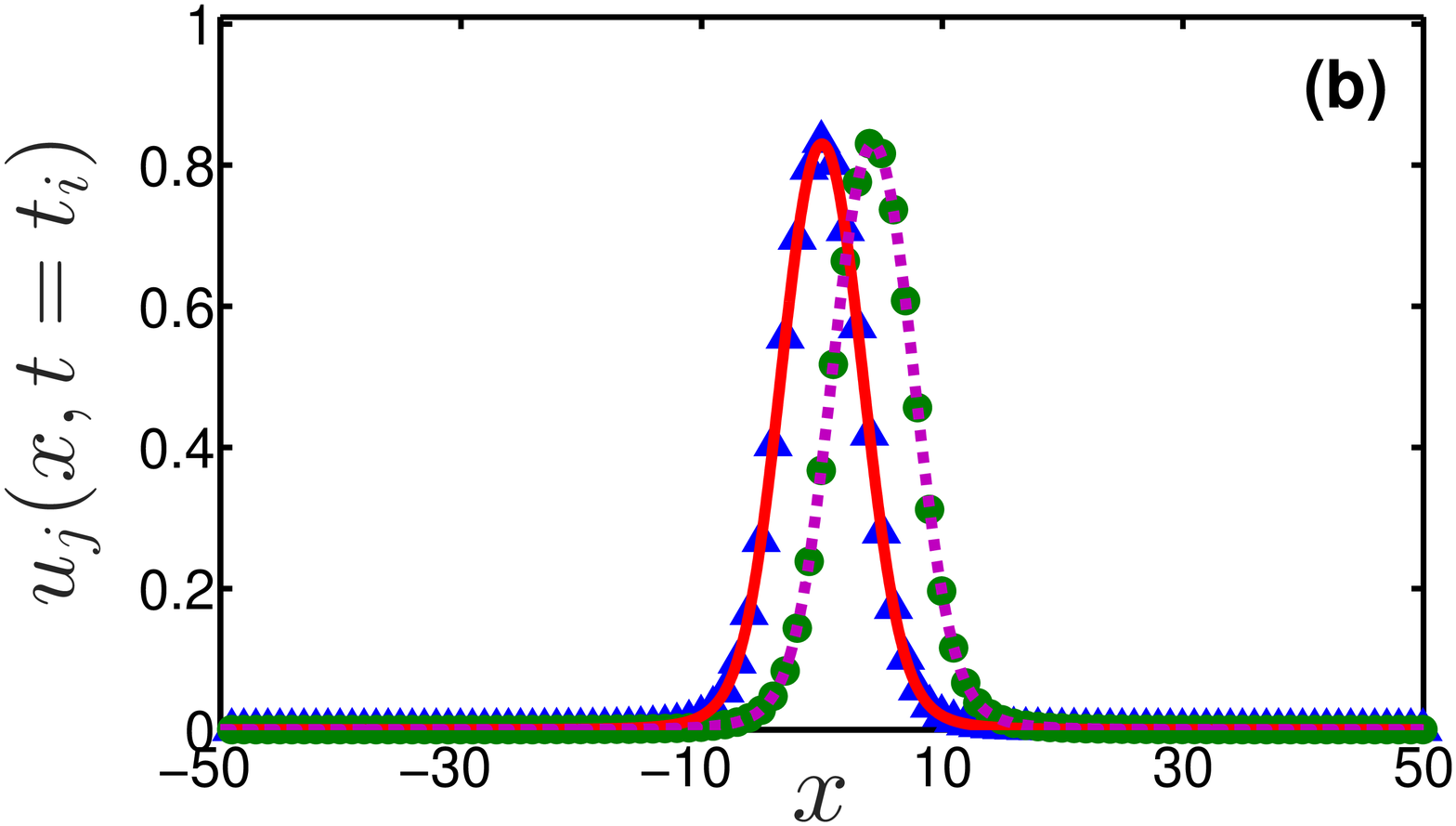} \\
\epsfxsize=8.5cm  \epsffile{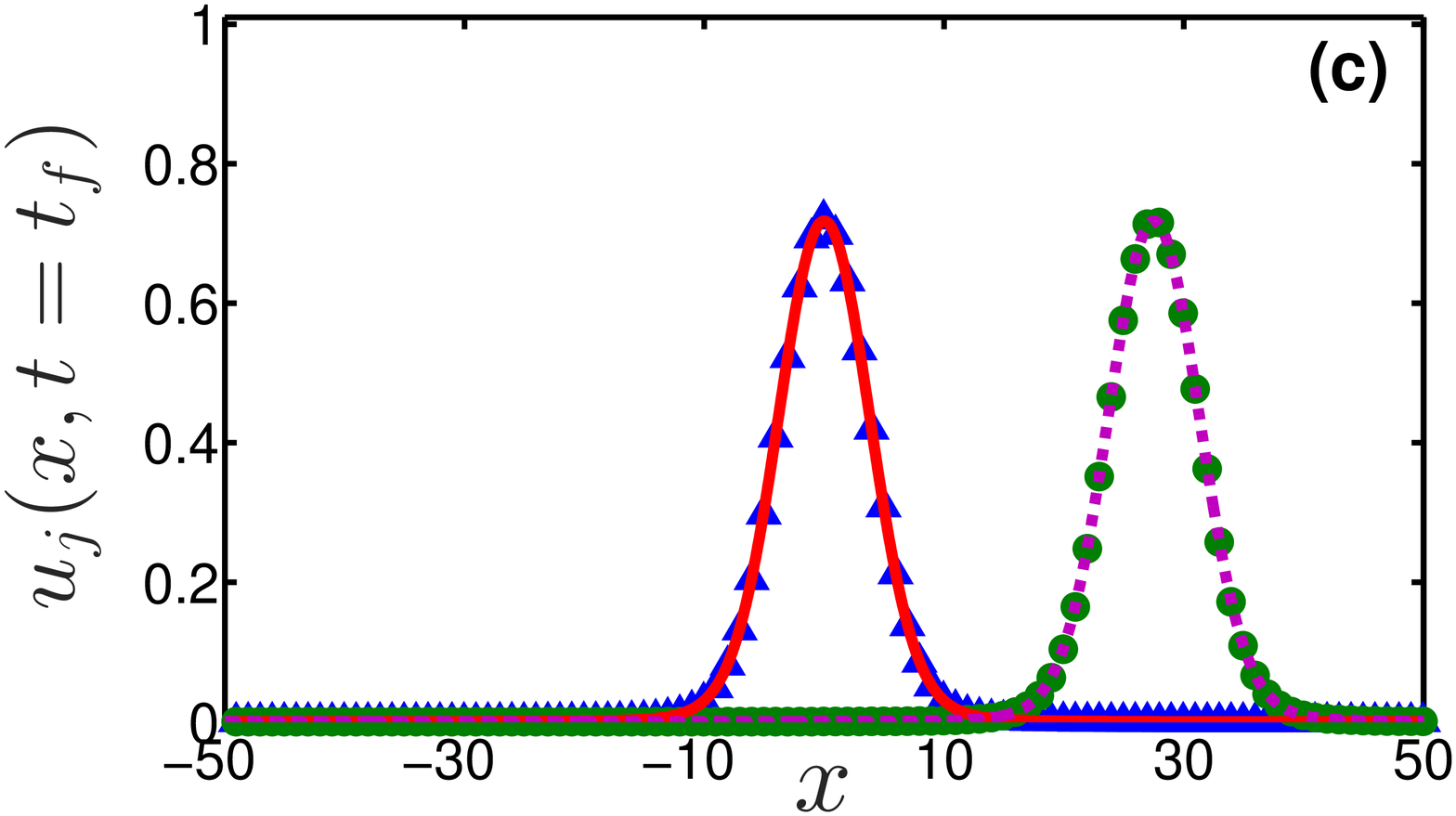} 
\end{tabular}
\caption{(Color online) The pulse shapes $u_j(x,t)$ at $t=0$ (a), $t=t_{i}=1.929$ (b), and $t=t_{f}=3.5$ (c) 
in a fast collision between two generalized Cauchy-Lorentz pulses in a system described by the 
coupled diffusion-advection model (\ref{rda1}). The advection velocity is $v_{d}=15$. 
The blue triangles and green circles represent the initial pulse shapes $u_j(x,0)$ with $j=1,2$ in (a), 
and the perturbation theory's prediction for $u_j(x,t)$ with $j=1,2$ in (b) and (c). 
The solid red and dashed magenta curves in (b) and (c) correspond to $u_j(x,t)$ with $j=1,2$, 
obtained by numerical solution of Eq. (\ref{rda1}).}
\label{fig9}
\end{figure}

\begin{figure}[ptb]
\epsfxsize=12.0cm  \epsffile{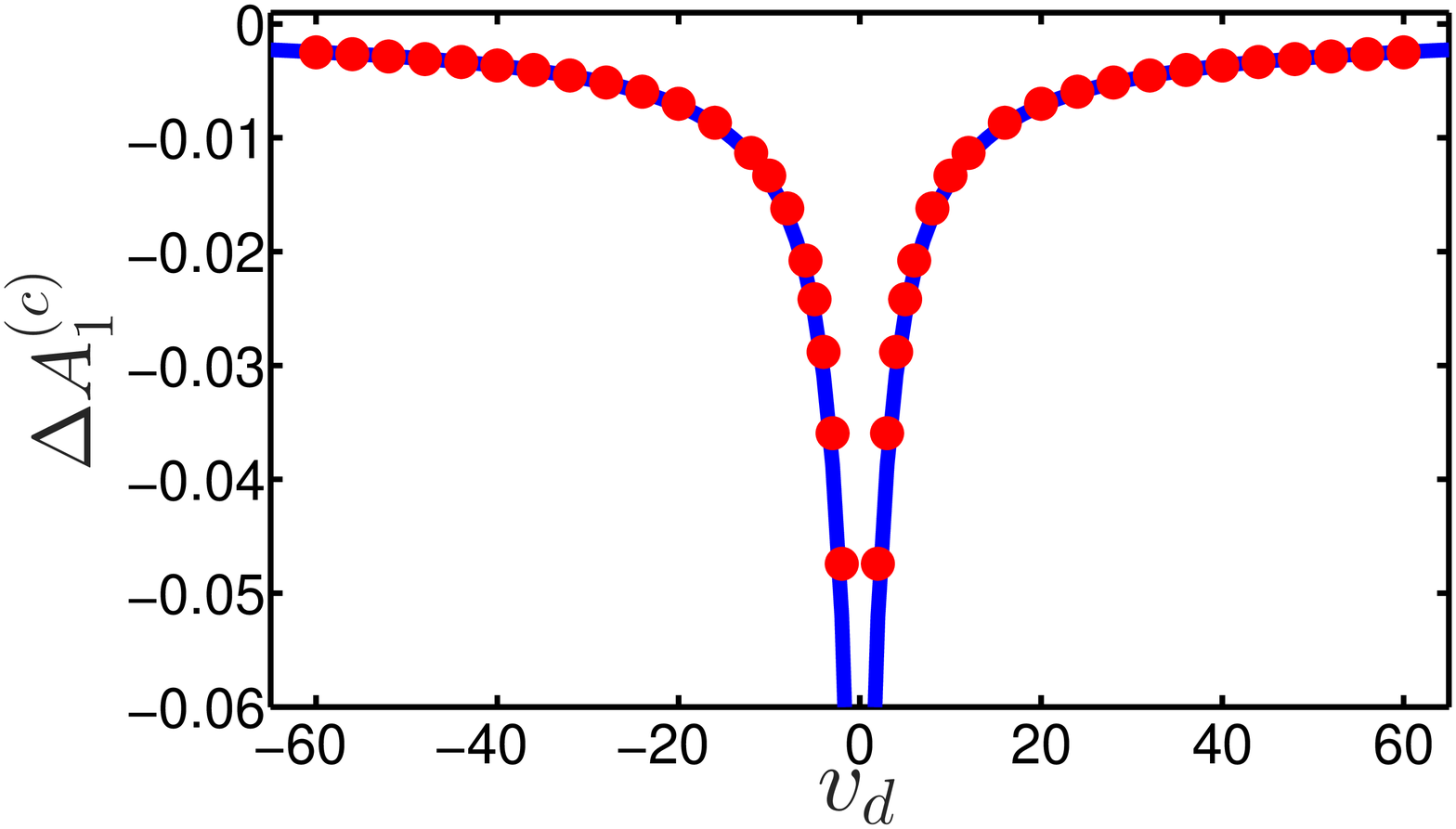} 
\caption{(Color online) The collision-induced amplitude shift of pulse 1 
$\Delta A_{1}^{(c)}$ vs advection velocity $v_{d}$ 
in a fast collision between two generalized Cauchy-Lorentz pulses in a system  
described by the coupled diffusion-advection model (\ref{rda1}).  
The red circles represent the result obtained by numerical solution of Eq. (\ref{rda1}). 
The solid blue curve corresponds to the prediction of Eq. (\ref{rda8}) with $C_{D}=2^{1/4}\pi$.}
\label{fig10}
\end{figure}

Next, we present the results of numerical simulations for fast collisions 
between generalized Cauchy-Lorentz pulses. 
Figure \ref{fig9} shows the pulse shapes $u_j(x,t)$ obtained in the simulation 
with $v_{d}=15$ at $t=0$, $t_{i}=1.929>t_{c}$, and $t_{f}=3.5$. 
Also shown is the analytic prediction for $u_j(x,t)$, which is obtained 
with Eq. (\ref{rda2}). The agreement between the numerical result 
and the analytic prediction is very good at both $t=t_{i}$ and $t=t_{f}$ 
despite of the diffusion-induced broadening experienced by the pulses. 
Additionally, we do not observe any noticeable oscillatory features in the pulse tails, 
such as the ones seen in Figs. \ref{fig3}(b) and \ref{fig3}(c) for collisions between 
generalized Cauchy-Lorentz pulses in linear optical waveguides. 
The dependence of $\Delta A_{1}^{(c)}$ on $v_{d}$ obtained in the simulations 
is shown in Fig. \ref{fig10} together with the analytic prediction of Eq. (\ref{rda8}). 
We observe very good agreement between the results of the simulations and the analytic 
prediction. Indeed, the relative error in the approximation of $\Delta A_{1}^{(c)}$        
is less than 3.3$\%$ for $10 \le |v_{d}| \le 60$ and less than 8.9$\%$ for $2 \le |v_{d}| < 10$.
Similar results are obtained for other values of the physical parameters 
and for other pulse shapes with power-law decreasing tails. We therefore 
conclude that the universal behavior of the collision-induced amplitude shift 
is also observed in fast collisions between pulses, whose tails exhibit relatively 
slow (power-law) decrease with increasing distance from the pulse maximum.                                                                              

\begin{figure}[ptb]
\begin{tabular}{cc}
\epsfxsize=8.5cm  \epsffile{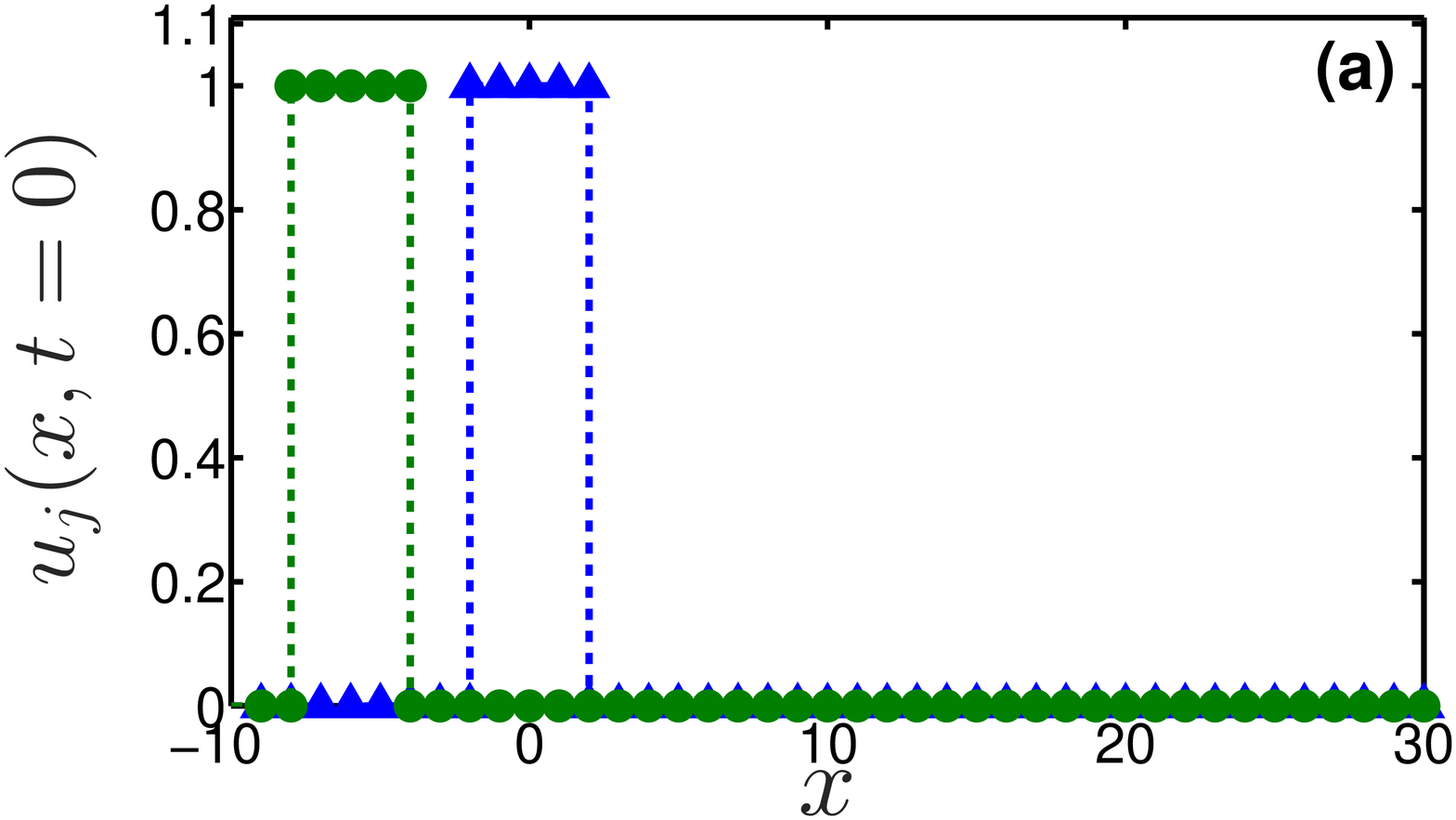} \\
\epsfxsize=8.5cm  \epsffile{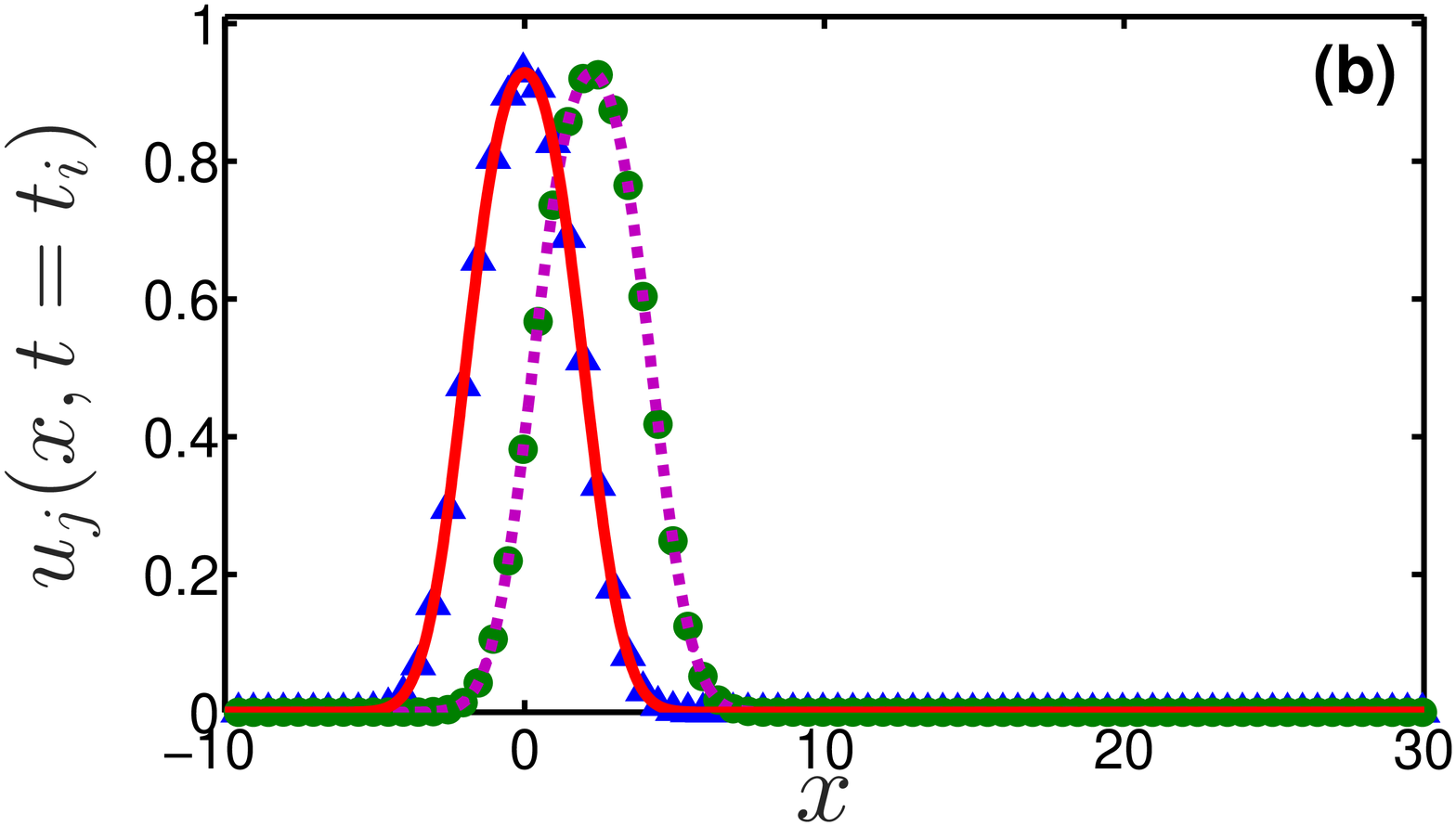} \\
\epsfxsize=8.5cm  \epsffile{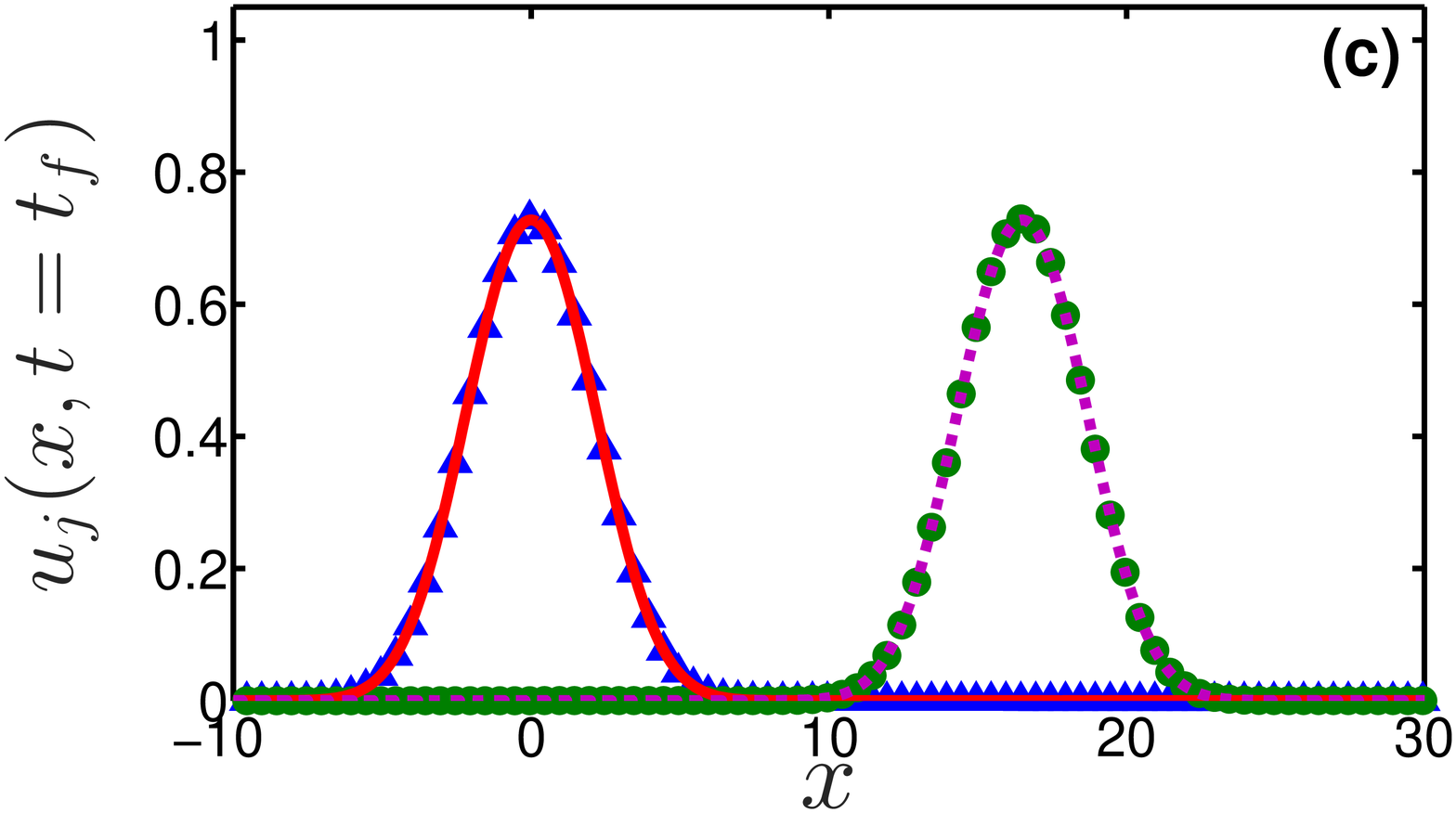} 
\end{tabular}
\caption{(Color online) The pulse shapes $u_j(x,t)$ at $t=0$ (a), $t=t_{i}=0.557$ (b), and $t=t_{f}=1.5$ (c) 
in a fast collision between two square pulses in a system described by the 
coupled diffusion-advection model (\ref{rda1}). The advection velocity is $v_{d}=15$. 
The blue triangles and green circles represent the initial pulse shapes $u_j(x,0)$ with $j=1,2$ in (a), 
and the perturbation theory's prediction for $u_j(x,t)$ with $j=1,2$ in (b) and (c). 
The solid red and dashed magenta curves in (b) and (c) correspond to $u_j(x,t)$ with $j=1,2$, 
obtained by numerical solution of Eq. (\ref{rda1}).}
\label{fig11}
\end{figure}

\begin{figure}[ptb]
\epsfxsize=12.0cm  \epsffile{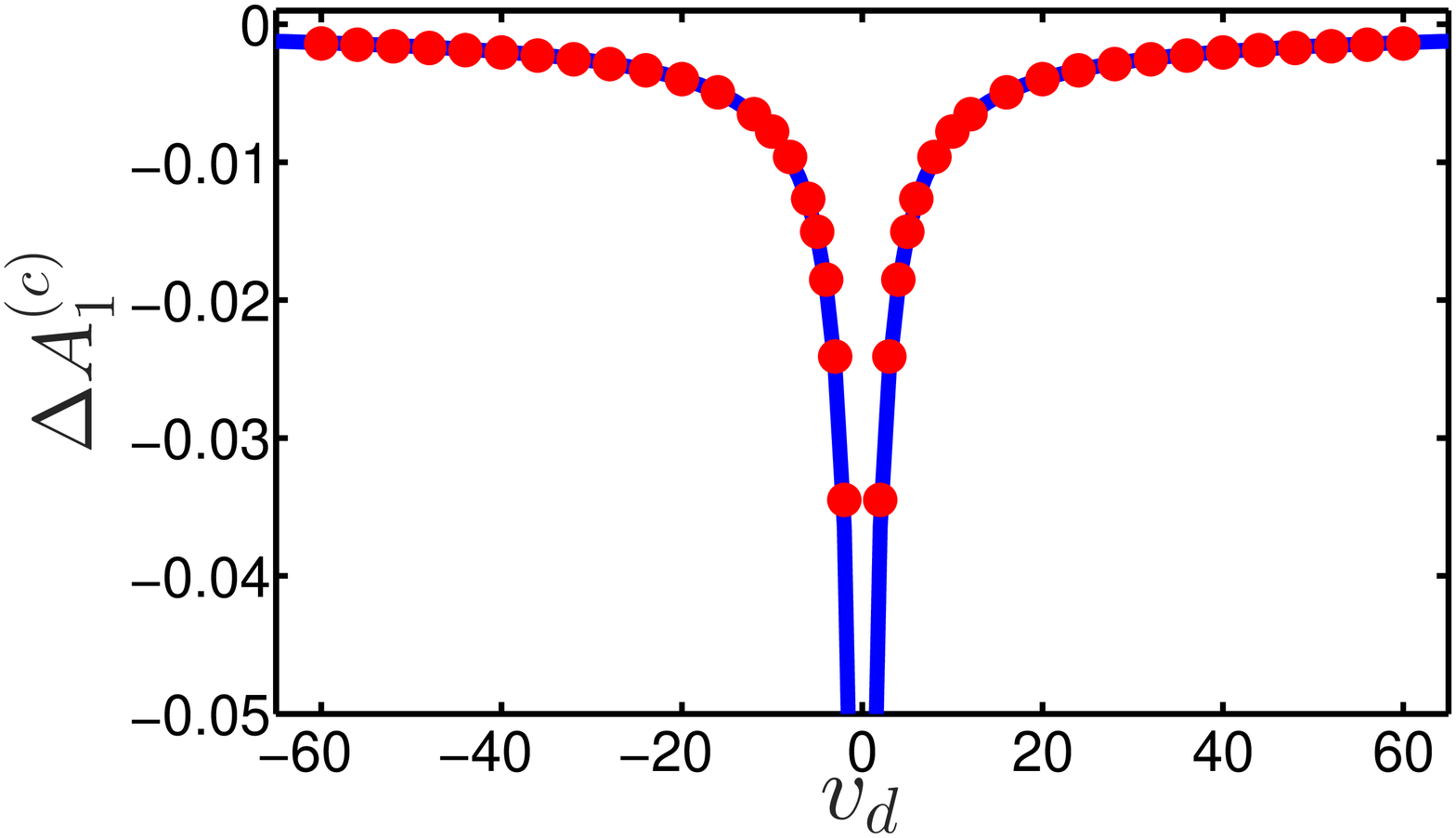} 
\caption{(Color online) The collision-induced amplitude shift of pulse 1 
$\Delta A_{1}^{(c)}$ vs advection velocity $v_{d}$ 
in a fast collision between two square pulses in a system  
described by the coupled diffusion-advection model (\ref{rda1}).  
The red circles represent the result obtained by numerical solution of Eq. (\ref{rda1}). 
The solid blue curve corresponds to the prediction of Eq. (\ref{rda8}) with $C_{D}=2$.}
\label{fig12}
\end{figure}

Finally, we describe the results of the simulations for fast collisions between square pulses. 
The initial pulse shapes $u_j(x,0)$, and the pulse shapes $u_j(x,t)$ obtained in the simulation 
with $v_{d}=15$ at $t=0$, $t_{i}=0.557>t_{c}$, and $t_{f}=1.5$ is shown in Fig. \ref{fig11}.  
The analytic prediction, obtained with Eq. (\ref{rda2}), is also shown. It is seen that the pulses 
undergo significant broadening due to the effects of diffusion. However, in contrast with the situation in 
linear optical waveguides, the pulses do not develop any observable oscillatory tails. 
Despite of the broadening, the agreement between the numerical result and the analytic prediction  
for the pulse shapes is very good. 
Figure \ref{fig12} shows the dependence of $\Delta A_{1}^{(c)}$ on $v_{d}$ 
obtained by the simulations together with the analytic prediction of Eq. (\ref{rda8}). 
We observe very good agreement between the analytic prediction and the simulations results. 
More specifically, the relative error in the approximation of $\Delta A_{1}^{(c)}$ 
is smaller than 4.3$\%$ for $10 \le |v_{d}| \le 60$ and smaller than 5.3$\%$ for $2 \le |v_{d}| < 10$.
Similar results are obtained with other values of the physical parameters. 
We therefore conclude that the universal behavior of the collision-induced amplitude shift 
can be observed even in collisions between pulses with nonsmooth initial shapes.

\section{Conclusions}
\label{conclusions}
We demonstrated that the amplitude shifts in fast two-pulse collisions 
in linear physical systems, weakly perturbed by nonlinear dissipation,  
exhibit universal soliton-like behavior. 
The behavior was demonstrated for linear optical waveguides 
with weak cubic loss and for systems described by 
linear diffusion-advection models with weak quadratic loss. 
We showed that in both cases, the expressions for the collision-induced 
amplitude shifts due to the nonlinear loss have the same form as the expression 
for the amplitude shift in a fast collision between two solitons 
of the cubic NLS equation in the presence of weak cubic loss.   
Furthermore, we showed that the expressions for the amplitude shifts are 
universal in the sense that they are independent  
of the exact details of the initial pulse shapes. 
The universal soliton-like behavior of the expressions for the 
collision-induced amplitude shifts was explained by noting that changes 
in pulse shapes occurring during the collision due to the effects of dispersion or diffusion 
can be neglected for fast collisions, 
and by noting the conservation of the total energies (or total masses) of the pulses  
by the unperturbed linear evolution models.   
We demonstrated the universal behavior of the collision-induced amplitude 
shifts by performing numerical simulations with the two perturbed 
coupled linear evolution models with three different initial conditions corresponding to 
pulses with exponentially decreasing tails, pulses with power-law decreasing tails, 
and pulses that are initially nonsmooth.   
In all six cases we found very good agreement between 
the analytic predictions for the amplitude shifts and the results of 
the numerical simulations.  
Surprisingly, the analytic predictions held even for collisions between  
pulses with initially nonsmooth shapes in linear optical waveguides despite 
of the fast generation of significant pulse tails. 
The good agreement between the analytic and numerical results in the latter 
case was explained by noting that during fast collisions most of the pulse energies 
are still contained in the main bodies of the pulses, and by noting the conservation 
of the total energies of the two pulses by the unperturbed linear propagation model.

\section*{Acknowledgments}
Q.M.N. and T.T.H. are supported by the Vietnam National Foundation 
for Science and Technology Development (NAFOSTED) 
under Grant No. 101.99-2015.29.

\appendix
\section{Calculation of $\Delta A_{1}^{(c)}$ from $\Delta\Phi_{1}(t,z_{c})$ 
and $\Delta\phi_{1}(x,t_{c})$}
\label{appendA}
In this Appendix, we derive relations (\ref{coll6_add1}) and (\ref{rda6_add1}) 
between the net collision-induced amplitude shift $\Delta A_{1}^{(c)}$ and the net 
collision-induced changes in the envelope of the electric field 
and in material concentration $\Delta\Phi_{1}(t,z_{c})$ and $\Delta\phi_{1}(x,t_{c})$. 
These relations were used to obtain Eq. (\ref{coll7}) and Eq. (\ref{rda7}) 
for $\Delta A_{1}^{(c)}$ from Eqs. (\ref{coll6}) and (\ref{rda6}), respectively.

We first consider the linear waveguide system with weak linear and 
cubic loss, described by Eq. (\ref{coll1}). 
To obtain the relation between $\Delta A_{1}^{(c)}$ and $\Delta\Phi_{1}(t,z_{c})$, 
we recall that the amplitude dynamics of a single pulse propagating in the presence 
of linear or nonlinear loss can be determined by an equation of the form 
$\partial_{z}\int_{-\infty}^{\infty} \!\!\!dt|\psi_{1}(t,z)|^{2}=...$, 
where the right hand side of the equation is determined by the character 
of the loss perturbation [see for example Eq. (\ref{app2_coll1}) is Appendix \ref{appendB}]. 
A fast collision that takes place at a distance $z=z_{c}$ 
leads to a jump in the value of $\int_{-\infty}^{\infty} \!\!\!dt|\psi_{1}(t,z)|^{2}$ 
at $z=z_{c}$. Therefore, in this case, the term $\partial_{z}\int_{-\infty}^{\infty} \!\!\!dt|\psi_{1}(t,z)|^{2}$ 
in the equation determining the dynamics of the amplitude should be replaced by the following quantity:
\begin{eqnarray}&&
\Delta_{P}=
\int_{-\infty}^{\infty} \!\!\!\! dt |\psi_{1}(t,z_{c}^{+})|^{2}-
\int_{-\infty}^{\infty} \!\!\!\! dt |\psi_{1}(t,z_{c}^{-})|^{2}.
\label{app1_collB0}
\end{eqnarray}      
The relation between $\Delta A_{1}^{(c)}$ and $\Delta\Phi_{1}(t,z_{c})$ is obtained 
by finding two expressions for $\Delta_{P}$, one involving $\Delta A_{1}^{(c)}$ 
and the other involving $\Delta\Phi_{1}(t,z_{c})$, and by equating the two expressions.

We note that by definition of $\psi_{10}$ and $\tilde\Psi_{10}$, 
$\psi_{1}(t,z_{c}^{-}) \simeq \psi_{10}(t,z_{c}^{-}) \simeq 
A_{1}(z_{c}^{-})\tilde\Psi_{10}(t,z_{c})\exp[i\chi_{10}(t,z_{c})]$.   
Therefore, we can write 
\begin{eqnarray}&&
\int_{-\infty}^{\infty} \!\!\!\! dt |\psi_{1}(t,z_{c}^{-})|^{2}= 
C_{1}A_{1}^{2}(z_{c}^{-}),
\label{app1_collB0_add1}
\end{eqnarray}          
where $C_{1}=\int_{-\infty}^{\infty} \!\!\! dt \tilde\Psi_{10}^{2}(t,z_{c})$ 
is a constant \cite{coll_conserved}. 
We also note that in the case of a fast collision, we can express $\Delta\phi_{1}(t,z_{c})$ as:   
$\Delta\phi_{1}(t,z_{c}) \simeq \phi_{1}(t,z_{c}^{+}) - \phi_{1}(t,z_{c}^{-}) \simeq \phi_{1}(t,z_{c}^{+})$. 
Using this relation along with Eq.  (\ref{coll2}) and the definition of $\psi_{10}$, we obtain:  
\begin{eqnarray}&&
\psi_{1}(t,z_{c}^{+})=\psi_{10}(t,z_{c}^{-})+\Delta\phi_{1}(t,z_{c}).
\label{app1_collB1}
\end{eqnarray}
Employing Eq. (\ref{app1_collB1}) together with the definitions of $\tilde\Psi_{10}$ 
and $\Delta\Phi_{1}$, we obtain:   
\begin{eqnarray}&&
\!\!\!\!\!\!\!\!\!\!\!\!
\int_{-\infty}^{\infty} \!\!\!\!\!\!\!\! dt |\psi_{1}(t,z_{c}^{+})|^{2}\!=\!\!\!
\int_{-\infty}^{\infty} \!\!\!\!\!\!\!\! dt \left[A_{1}(z_{c}^{-})\tilde\Psi_{10}(t,z_{c})\!
+\!\Delta\Phi_{1}(t,z_{c})\right]^{2} 
\!\!\!\!. 
%\,
\label{app1_collB2}
\end{eqnarray}     
Expanding the integrand on the right hand side of Eq.  (\ref{app1_collB2}), 
while keeping only the first two leading terms, we arrive at: 
\begin{eqnarray}&&
\!\!\!\!\!\!\!\!\!
\int_{-\infty}^{\infty} \!\!\!\!\!\!\!\! dt |\psi_{1}(t,z_{c}^{+})|^{2}\simeq
C_{1}A_{1}^{2}(z_{c}^{-})
\nonumber \\&&
+2A_{1}(z_{c}^{-})
\int_{-\infty}^{\infty} \!\!\!\!\!\!\!\! dt \tilde\Psi_{10}(t,z_{c})\Delta\Phi_{1}(t,z_{c}).
\label{app1_collB3}
\end{eqnarray}      
Substituting Eqs. (\ref{app1_collB0_add1}) and (\ref{app1_collB3}) into Eq. (\ref{app1_collB0}), 
we obtain the first expression for $\Delta_{P}$: 
\begin{eqnarray}&&
\Delta_{P}=
2A_{1}(z_{c}^{-})
\int_{-\infty}^{\infty} \!\!\!\!\!\!\!\! dt \tilde\Psi_{10}(t,z_{c})\Delta\Phi_{1}(t,z_{c}).
\label{app1_collB3_add1}
\end{eqnarray}        
On the other hand, we can express $\int_{-\infty}^{\infty} \!\!\! dt |\psi_{1}(t,z_{c}^{+})|^{2}$ 
in terms of $\Delta A_{1}^{(c)}$ in the following manner:  
\begin{eqnarray}&&
\!\!\!\!\!\!\!\!\!
\int_{-\infty}^{\infty} \!\!\!\!\!\!\!\! dt |\psi_{1}(t,z_{c}^{+})|^{2}\!=\!
\left(A_{1}(z_{c}^{-}) +\Delta A_{1}^{(c)}\right)^{2}
\int_{-\infty}^{\infty} \!\!\!\!\!\!\!\! dt \tilde\Psi_{10}^{2}(t,z_{c})
\nonumber \\&&
\simeq
C_{1}A_{1}^{2}(z_{c}^{-})+2C_{1}A_{1}(z_{c}^{-}) \Delta A_{1}^{(c)}.
\label{app1_collB4}
\end{eqnarray}
Substituting Eqs. (\ref{app1_collB0_add1}) and (\ref{app1_collB4}) into Eq. (\ref{app1_collB0}), 
we find the second expression for $\Delta_{P}$: 
\begin{eqnarray}&&
\Delta_{P}=2C_{1}A_{1}(z_{c}^{-}) \Delta A_{1}^{(c)}.
\label{app1_collB4_add1}
\end{eqnarray}                            
Equating the right hand sides of Eqs. (\ref{app1_collB3_add1}) and (\ref{app1_collB4_add1}), we obtain: 
\begin{eqnarray}&&
\!\!\!\!\!\!\!\!\!
\Delta A_{1}^{(c)}=
\frac{1}{C_{1}}
\int_{-\infty}^{\infty} \!\!\!\!\!\!\!\! dt \tilde\Psi_{10}(t,z_{c})\Delta\Phi_{1}(t,z_{c}), 
\label{app1_collB5}
\end{eqnarray}
which is  Eq. (\ref{coll6_add1}).

We now treat systems described by the coupled linear diffusion-advection model (\ref{rda1}). 
The dynamics of the amplitude of a single pulse propagating in the presence 
of linear or nonlinear loss can be determined by an equation of the form 
$\partial_{t}\int_{-\infty}^{\infty} \!\!\!dx u_{1}(x,t)=...$, 
where the right hand side of the equation is determined by the nature
of the loss perturbation. A fast collision that takes place at time $t=t_{c}$ 
leads to a jump in the value of $\int_{-\infty}^{\infty} \!\!\!dx u_{1}(x,t)$ at $t=t_{c}$. 
Therefore, in the fast collision problem, the term $\partial_{t}\int_{-\infty}^{\infty} \!\!\!dx u_{1}(x,t)$  
in the equation that determines the dynamics of the pulse amplitude should be replaced by:
\begin{eqnarray}&&
\Delta_{D}=
\int_{-\infty}^{\infty} \!\!\!\!\!\!\!\! dx \, u_{1}(x,t_{c}^{+})-
\int_{-\infty}^{\infty} \!\!\!\!\!\!\!\! dx \, u_{1}(x,t_{c}^{-}).
\label{app1_collB6_add1}
\end{eqnarray}      
We now find two expressions for $\Delta_{D}$, one that depends on $\Delta A_{1}^{(c)}$ 
and the other that depends on $\Delta\phi_{1}(x,t_{c})$. The relation between 
$\Delta A_{1}^{(c)}$ and $\Delta\phi_{1}(x,t_{c})$ is obtained by 
equating the two expressions.

By the definitions of $u_{1}$, $u_{10}$, and $\tilde u_{10}$, 
$u_{1}(x,t_{c}^{-}) \simeq u_{10}(x,t_{c}^{-}) \simeq A_{1}(t_{c}^{-}) \tilde u_{10}(x,t_{c})$. 
It follows that 
\begin{eqnarray}&&
\int_{-\infty}^{\infty} \!\!\!\!\!\! dx \, u_{1}(x,t_{c}^{-}) = C_{2}A_{1}(t_{c}^{-}),
\label{app1_collB6_add2}
\end{eqnarray}       
where $C_{2}=\int_{-\infty}^{\infty} \!\!\!\!\! dx \tilde u_{10}(x,t_{c})$ 
is a constant \cite{rda_conserved}. In addition, in a fast collision, 
we can express $\Delta\phi_{1}(x,t_{c})$ as: 
$\Delta\phi_{1}(x,t_{c}) \simeq \phi_{1}(x,t_{c}^{+}) - \phi_{1}(x,t_{c}^{-})  \simeq \phi_{1}(x,t_{c}^{+})$. 
Using this relation together with Eq. (\ref{rda2}) and the definition of $u_{10}$, we obtain  
\begin{eqnarray}&&
u_{1}(x,t_{c}^{+})=u_{10}(x,t_{c}^{-})+\Delta\phi_{1}(x,t_{c}).
\label{app1_collB6_add3}
\end{eqnarray}       
From Eq.  (\ref{app1_collB6_add3}), it follows that 
\begin{eqnarray}&&
\!\!\!\!\!\!\!\!\!\!\!\!\!\!\!\!
\int_{-\infty}^{\infty} \!\!\!\!\!\!\!\! dx \, u_{1}(x,t_{c}^{+})=
\int_{-\infty}^{\infty} \!\!\!\!\!\!\!\! dx 
\left[u_{10}(x,t_{c}^{-})+\Delta\phi_{1}(x,t_{c})\right].
\label{app1_collB6}
\end{eqnarray}  
Using the definition of $\tilde u_{10}$ in Eq. (\ref{app1_collB6}),  
we arrive at 
\begin{eqnarray}&&
\!\!\!\!\!\!\!\!\!\!\!\!\!\!\!\!
\int_{-\infty}^{\infty} \!\!\!\!\!\!\!\! dx \, u_{1}(x,t_{c}^{+})=C_{2}A_{1}(t_{c}^{-})+
\int_{-\infty}^{\infty} \!\!\!\!\!\!\!\! dx \, \Delta\phi_{1}(x,t_{c}).
\label{app1_collB7}
\end{eqnarray}  
Substituting Eqs. (\ref{app1_collB6_add2}) and (\ref{app1_collB7}) into Eq. (\ref{app1_collB6_add1}), 
we obtain the first expression for $\Delta_{D}$: 
\begin{eqnarray}&&
\Delta_{D}=
\int_{-\infty}^{\infty} \!\!\!\!\!\!\!\! dx \, \Delta\phi_{1}(x,t_{c}).
\label{app1_collB7_add1}
\end{eqnarray}  
On the other hand, we can express $\int_{-\infty}^{\infty} \!\!\! dx \, u_{1}(x,t_{c}^{+})$ 
in terms of $\Delta A_{1}^{(c)}$ in the following way:  
\begin{eqnarray}&&
\!\!\!\!\!\!\!\!\!\!\!\!\!\!\!\!
\int_{-\infty}^{\infty} \!\!\!\!\!\!\!\! dx \, u_{1}(x,t_{c}^{+})=
\left(A_{1}(t_{c}^{-})+\Delta A_{1}^{(c)}\right)
\int_{-\infty}^{\infty} \!\!\!\!\!\!\!\! dx \, \tilde u_{10}(x,t_{c})
\nonumber \\&&
=C_{2}A_{1}(t_{c}^{-})+C_{2} \Delta A_{1}^{(c)}.
\label{app1_collB8}
\end{eqnarray}  
Substituting Eqs. (\ref{app1_collB6_add2}) and (\ref{app1_collB8}) into Eq. (\ref{app1_collB6_add1}), 
we obtain the second expression for $\Delta_{D}$: 
\begin{eqnarray}&&
\Delta_{D} = C_{2} \Delta A_{1}^{(c)}.
\label{app1_collB8_add1}
\end{eqnarray}  
Equating the right hand sides of Eqs. (\ref{app1_collB7_add1}) and (\ref{app1_collB8_add1}), 
we obtain: 
\begin{eqnarray}&&
\!\!\!\!\!\!\!\!\!
\Delta A_{1}^{(c)}=
\frac{1}{C_{2}}
\int_{-\infty}^{\infty} \!\!\!\!\!\!\!\! dx \, \Delta\phi_{1}(x,t_{c}),
\label{supp_collB9}
\end{eqnarray}  
which is Eq. (\ref{rda6_add1}).

\section{Methods for calculating the values of $\Delta A_{1}^{(c)}$
from the analytic predictions and from numerical simulations}
\label{appendB}
In this appendix, we describe the methods used for obtaining the 
values of the collision-induced amplitude shift $\Delta A_{1}^{(c)}$
from the analytic predictions and from results of numerical simulations. 
We demonstrate the implementation of these methods for a collision between 
pulses with generic shapes in a linear optical waveguide with weak linear and cubic loss. 
The implementation of the methods for collisions in physical systems described by linear 
diffusion-advection models is similar.

The analytic prediction for $\Delta A_{1}^{(c)}$ is obtained by 
employing Eq. (\ref{coll7_add1}). The values of $A_{j}(z_{c}^{-})$ 
appearing in this equation are calculated by solving approximate equations  
for the dynamics of the $A_{j}(z)$ for a single pulse, 
propagating in the presence of first and second-order dispersion, 
linear loss, and cubic loss. More specifically, single-pulse propagation of pulse 1 
is described by Eq. (\ref{coll2_add1}) and single-pulse propagation of pulse 2 
is described by Eq. (\ref{coll2_add2}). Employing energy balance calculations 
for these two propagation models, we obtain
\begin{eqnarray}&&
\!\!\!\!\!\!\!\!\!\!\!\!
\partial_{z}\int_{-\infty}^{\infty} \!\!\!\!\!\!dt|\psi_{j0}|^{2}\!=\!
-2\epsilon_{1}\int_{-\infty}^{\infty} \!\!\!\!\!\!dt|\psi_{j0}|^{2}
-2\epsilon_{3}\int_{-\infty}^{\infty} \!\!\!\!\!\!dt|\psi_{j0}|^{4}.
\label{app2_coll1}
\end{eqnarray}       
We express the approximate solutions of Eq. (\ref{coll2_add1}) and (\ref{coll2_add2}) 
as $\psi_{j0}(t,z)=A_{j}(z)\tilde\psi_{j0}(t,z)$,  
where $\tilde\psi_{j0}(t,z)=\tilde\Psi_{j0}(t,z)\exp[i\chi_{j0}(t,z)]$ 
is the solution of the unperturbed linear propagation equation with    
initial amplitude $A_{j}(0)=1$. Substituting the approximate expressions for 
$\psi_{j0}(t,z)$ into Eq. (\ref{app2_coll1}), we obtain: 
\begin{eqnarray}&&
\!\!\!\!\!\!\!\!\!\!\!\!
\frac{d}{dz} 
\left[I_{2j}A_{j}^{2}\right]= 
-2\epsilon_{1}I_{2j}A_{j}^{2}-2\epsilon_{3}I_{4j}(z)A_{j}^{4} \,,  
\label{app2_coll2}
\end{eqnarray}
where $I_{2j}=\int_{-\infty}^{\infty} \!\!\!dt \tilde\Psi_{j0}^{2}(t,z)=
\int_{-\infty}^{\infty} \!\!\!dt \tilde\Psi_{j0}^{2}(t,0)=\mbox{const}$ and 
$I_{4j}(z)=\int_{-\infty}^{\infty} \!\!\!dt \tilde\Psi_{j0}^{4}(t,z)$.  
Equation (\ref{app2_coll2}) is a Bernoulli equation for $A_{j}^{2}(z)$. 
Its solution on the interval $[0,z_{c}]$ yields 
the following expression for $A_{j}(z_{c}^{-})$: 
\begin{eqnarray}&&
\!\!\!\!\!\!\!\!\!\!\!\!
A_{j}(z_{c}^{-})=\frac{A_{j}(0)e^{-\epsilon_{1}z_{c}}}
{\left[1+2\epsilon_{3}\tilde I_{4j}(0,z_{c})A_{j}^{2}(0)/I_{2j}\right]^{1/2}} \,,  
\label{app2_coll4}
\end{eqnarray}  
where 
\begin{eqnarray}&&
\!\!\!\!\!\!\!\!\!\!\!\!
\tilde I_{4j}(z_{1},z_{2})=\int_{z_{1}}^{z_{2}}
dz \, I_{4j}(z) e^{-2\epsilon_{1}z} \,.
\label{app2_coll7}
\end{eqnarray}      
We obtain the analytic prediction for $\Delta A_{1}^{(c)}$ by calculating 
the values of $A_{j}(z_{c}^{-})$ with Eq. (\ref{app2_coll4}) and by substituting 
these values into Eq. (\ref{coll7_add1}).

We obtain the value of $\Delta A_{1}^{(c)}$ from 
the results of the numerical simulations by using the relation 
\begin{eqnarray}&&
\Delta A_{1}^{(c)}=A_{1}(z_{c}^{+})-A_{1}(z_{c}^{-}) \,,
\label{app2_coll8}
\end{eqnarray}       
where $A_{1}(z_{c}^{-})$ is calculated with Eq. (\ref{app2_coll4}),  
and $A_{1}(z_{c}^{+})$ is determined from the simulations. 
More specifically, we solve Eq. (\ref{app2_coll2}) 
with $j=1$ on the interval $[z_{c},z_{f}]$ and obtain 
\begin{eqnarray}&&
\!\!\!\!\!\!\!\!\!\!\!\!
A_{1}(z_{c}^{+})=\frac{A_{1}(z_{f})e^{-\epsilon_{1}z_{c}}}
{\left[e^{-2\epsilon_{1}z_{f}}
-2\epsilon_{3}\tilde I_{41}(z_{c},z_{f})A_{1}^{2}(z_{f})/I_{21}\right]^{1/2}} \,.  
\label{app2_coll9}
\end{eqnarray}    
We then determine the value of $A_{1}(z_{c}^{+})$ by using Eq. (\ref{app2_coll9})   
with a value of $A_{1}(z_{f})$, which is measured from the simulations.


\begin{thebibliography}{}
\bibitem{fast_collisions} Fast soliton collisions are collisions for which 
the difference between the central frequencies (or group velocities) 
of the solitons is much larger than the soliton spectral width.
\bibitem{Dimensions1} The dimensionless distance $z$ in Eq. (\ref{coll1})  is 
$z=Z/(2L_{D})$, where $Z$ is the dimensional distance, 
$L_{D}=\tau_{0}^{2}/|\tilde\beta_{2}|$ is the dispersion length,
and $\tau_{0}$ is a reference pulse width. 
The dimensionless time is $t=\tau/\tau_{0}$, where $\tau$ is time. 
$\psi_{j}=E_{j}/\sqrt{P_{0}}$, where $E_{j}$ is the 
electric field of the $j$th pulse and $P_{0}$ is peak power.   
$d_{1}=2(\tilde\beta_{12}-\tilde\beta_{11})\tau_{0}/|\tilde\beta_{2}|$, 
where $\tilde\beta_{1j}=n_{gj}/c=1/v_{gj}$, $c$ is the speed of light,  
and $n_{gj}$ and $v_{gj}$ are the group refractive index 
and the group velocity for the $j$th pulse. 
$\epsilon_{1}=2\tau_{0}^{2}\tilde\rho_{1}/|\tilde\beta_{2}|$  
and $\epsilon_{3}=2P_{0}\tau_{0}^{2}\tilde\rho_{3}/|\tilde\beta_{2}|$, 
where $\tilde\rho_{1}$ and $\tilde\rho_{3}$ are  
the dimensional linear and cubic loss coefficients.
\bibitem{zi_values} The values of $z_{i}$ are determined by the equation 
$z_{i}=z_{c}+r(z_{f}-z_{c})$, where $z_{c}=(y_{10}-y_{20})/d_{1}$, 
and $r=1/7$, as an example. Thus, $z_{i}$ is an intermediate distance 
that is larger than $z_{c}$, at which the collision is not yet completed.
\bibitem{Dimensions2} The dimensionless coordinate $x$ in Eq. (\ref{rda1}) is 
 $x=X/x_{0}$, where $X$ is the dimensional coordinate, and $x_{0}$ is a reference pulse width.  
The dimensionless time is $t=\tau/\tau_{D}$, where $\tau$ is time, $\tau_{D}=x_{0}^{2}/D$, 
and $D$ is the diffusion coefficient.     
$u_{j}=U_{j}/\rho_{0}$, where $U_{j}$ is the 
concentration of substance $j$ and $\rho_{0}$ is the peak concentration.  
$v_{d}=x_{0}V_{d}/D$, where $V_{d}$ is the dimensional advection velocity.  
$\varepsilon_{1}=x_{0}^{2}\tilde\varepsilon_{1}/D$ and   
$\varepsilon_{2}=\rho_{0}x_{0}^{2}\tilde\varepsilon_{2}/D$, 
where $\tilde\varepsilon_{1}$ and $\tilde\varepsilon_{2}$ are the  
dimensional linear and quadratic loss coefficients.
\bibitem{ti_values} The values of $t_{i}$ are determined by the equation 
$t_{i}=t_{c}+r(t_{f}-t_{c})$, where $t_{c}=(x_{10}-x_{20})/v_{d}$, 
and $r=1/7$, as an example. Thus, $t_{i}$ is an intermediate time  
that is larger than $t_{c}$, at which the collision is not yet completed.
\bibitem{coll_conserved} Since the integral $\int_{-\infty}^{\infty} \!\!\! dt \tilde\Psi_{10}^{2}(t,z)$ 
is conserved by the unperturbed linear propagation equation, we can write: 
$C_{1}=\int_{-\infty}^{\infty} \!\!\! dt \tilde\Psi_{10}^{2}(t,0)$.
\bibitem{rda_conserved}
Since the integral $\int_{-\infty}^{\infty} \!\!\! dx \tilde u_{10}(x,t)$ 
is conserved by the unperturbed linear diffusion equation, we can write: 
$C_{2}=\int_{-\infty}^{\infty} \!\!\! dx \tilde u_{10}(x,0)$.
\end{thebibliography}
\end{document}